\documentclass[acmsmall]{acmart}
\AtBeginDocument{%
  }

\acmJournal{TOSEM}
\setcopyright{none}
\acmDOI{}
\acmVolume{}
\acmNumber{}

\makeatletter
\AtBeginDocument{%
  \ifnum\acm@copyrightmode=0\relax
    \let\footnotetextcopyrightpermission\@gobble
    \patchcmd{\@mkbibcitation}{\egroup. \@acmYear. \@title}
      {\egroup. \@title}{}{}%
    \patchcmd{\@mkbibcitation}
      {\@acmVolume, \@acmNumber \@article@string (\@acmPubDate),}
      {\unskip,}{}{}%
    \apptocmd{\ps@standardpagestyle}{%
      \fancyfoot[RO,LE]{\footnotesize \@journalNameShort}%
    }{}{}%
    \apptocmd{\ps@firstpagestyle}{%
      \fancyfoot[RO,LE]{\footnotesize \@journalNameShort}%
    }{}{}%
    \pagestyle{standardpagestyle}%
  \fi
}
\makeatother

\usepackage{color,soul}
\definecolor{revisionblue}{RGB}{125,211,252}
\sethlcolor{revisionblue}
\usepackage{multirow}
\usepackage{threeparttable}
\usepackage{booktabs}
\usepackage{subfigure}
\usepackage[many]{tcolorbox} 
\usepackage{fontawesome5} 
\usepackage{colortbl}
\usepackage{enumitem}
\usepackage{threeparttable}
\usepackage{placeins}
\usepackage{soul}
\soulregister\cite7
\soulregister\ref7
\usepackage[ruled,vlined,linesnumbered]{algorithm2e}
\SetAlgoVlined 
\SetAlgoNoEnd  
\DontPrintSemicolon 
\SetKwInput{KwInput}{Input}
\SetKwInput{KwOutput}{Output}
\newtcolorbox{boxK}{
    top=2pt,
    bottom=2pt,
    left=2pt,
    right=2pt,
    boxrule = 0pt,
    toprule = 0pt, 
}

\begin{document}

\title{DebugRepair: Enhancing LLM-Based Automated Program Repair via Self-Directed Debugging}

\author{Linhao Wu}
\authornote{Both authors contributed equally to this research.}
\email{wulinhao26@stu.pku.edu.cn}
\orcid{0009-0001-7624-156X}
\affiliation{%
  \institution{Shandong University}
  \city{Qingdao}
  \country{China}
}

\author{Yifei Pei}
\authornotemark[1]
\email{peiyifei@mail.sdu.edu.cn}
\orcid{0009-0005-1351-0565}
\affiliation{%
  \institution{Shandong University}
  \city{Qingdao}
  \country{China}
}

\author{Zhen Yang}
\authornote{Corresponding author.}
\email{zhenyang@sdu.edu.cn}
\orcid{0000-0003-0670-4538}
\affiliation{%
  \institution{Shandong University}
  \city{Qingdao}
  \country{China}
}

\author{Kainan Li}
\email{likainan@mail.sdu.edu.cn}
\orcid{0009-0008-1889-3843}
\affiliation{%
  \institution{Shandong University}
  \city{Qingdao}
  \country{China}
}

\author{Zhonghang Lu}
\email{25213050288@m.fudan.edu.cn}
\orcid{0009-0008-2193-2540}
\affiliation{%
  \institution{Fudan University}
  \city{Shanghai}
  \country{China}
}

\author{Hao Tan}
\email{202400130169@mail.sdu.edu.cn}
\orcid{0009-0007-4707-5113}
\affiliation{%
  \institution{Shandong University}
  \city{Qingdao}
  \country{China}
}

\author{Xiran Lyu}
\email{xiranlyu@mail.sdu.edu.cn}
\orcid{0009-0002-8861-8907}
\affiliation{%
  \institution{Shandong University}
  \city{Qingdao}
  \country{China}
}

\author{Jia Li}
\email{lijia@stu.pku.edu.cn}
\orcid{0000-0002-5579-8852}
\affiliation{%
  \institution{Tsinghua University}
  \city{Beijing}
  \country{China}
}

\author{Yizhou Chen}
\email{yizhouchen@stu.pku.edu.cn}
\orcid{0000-0003-1821-3170}
\affiliation{%
  \institution{Peking University}
  \city{Beijing}
  \country{China}
}

\author{Pengyu Xue}
\email{xuepengyu@mail.sdu.edu.cn}
\orcid{0009-0007-3395-9575}
\affiliation{%
  \institution{The Hong Kong Polytechnic University}
  \city{Hong Kong}
  \country{China}
}

\author{Kunwu Zheng}
\email{xiaozhengsdu2022@mail.sdu.edu.cn}
\orcid{0009-0000-7047-1981}
\affiliation{%
  \institution{Shandong University}
  \city{Qingdao}
  \country{China}
}

\author{Dan Hao}
\email{haodan@pku.edu.cn}
\orcid{0000-0001-8295-303X}
\affiliation{%
  \institution{Peking University}
  \city{Beijing}
  \country{China}
}

\renewcommand{\shortauthors}{Wu et al.}

\begin{abstract}
Automated Program Repair (APR) has recently benefited from the strong code understanding and generation capabilities of Large Language Models (LLMs). Among existing LLM-based APR techniques, feedback-based approaches have shown promising results by iteratively refining candidate patches according to test execution feedback. However, most of these approaches rely primarily on outcome-level failure symptoms, e.g., stack traces, which indicate how failures are observed but fail to expose the intermediate runtime states that are often critical for root-cause analysis. As a result, LLMs must infer bug causes without access to such intermediate runtime evidence, often leading to incorrect patches.

To address this limitation, we propose DebugRepair, a self-directed debugging framework for LLM-based APR. The key idea of DebugRepair is to enhance patch refinement with intermediate runtime evidence collected through simulated debugging, rather than relying solely on outcome-level failure symptoms. Specifically, DebugRepair consists of three components. \textbf{First}, test semantic purification extracts the minimal failure-triggering test context, thereby removing noise in both tests and follow-up debugging logs. \textbf{Second}, simulated instrumentation enables the LLM to insert targeted debugging statements into buggy functions to collect runtime traces. A rule-based fallback mechanism is invoked when LLM-instrumented code fails to compile or semantically introduces inconsistencies.
\textbf{Third}, debugging-driven conversational repair organizes patch generation into a hierarchical, iterative process, in which the LLM progressively refines candidate patches using both prior repair attempts and newly observed runtime states. 

{We evaluate DebugRepair on four widely used benchmarks across two Programming Languages (PLs), e.g., Java and Python.} Extensive experiments demonstrate the State-Of-The-Art (SOTA) performance of DebugRepair against 15 representative approaches.  For example, with GPT-3.5 as the backbone model, DebugRepair correctly fixes 224 bugs on Defects4J, achieving an average improvement of 26.2\% over SOTA LLM-based approaches. With DeepSeek-V3, DebugRepair correctly fixes 295 Defects4J bugs, exceeding the second-best baseline by 59 correct fixes. {Across six backbone LLMs of different families and sizes, DebugRepair improves repair performance by 48.4\% on average over their vanilla settings, demonstrating its model-agnostic effectiveness.} Further ablation studies confirm that all components of DebugRepair contribute effectively to the overall repair performance.

\end{abstract}



\maketitle

\section{Introduction}


{Automated Program Repair (APR) aims to automatically generate patches to fix software bugs \cite{gazzola2018automatic, le2019automated, zhang2023survey}. 
Existing APR techniques can be broadly grouped into three categories: traditional APR, learning-based APR, and LLM-based APR. 
Traditional APR techniques can be further categorized into three paradigms: template-based \cite{martinez2016astor, hua2018sketchfix, ghanbari2019practical, liu2019avatar}, heuristic-based \cite{le2011genprog, jiang2018shaping, le2016history, wen2018context}, and constraint-based \cite{long2015staged, le2017s3, mechtaev2016angelix, gao2021beyond} approaches. 
With advances in deep learning, learning-based APR emerged as another major line of research, demonstrating superior generality by learning code repair patterns from large-scale corpora. Learning-based approaches can be primarily divided into two categories: Neural Machine Translation (NMT)-based approaches and Pre-trained Language Model (PLM)-based approaches. The former \cite{chen2019sequencer, jiang2021cure, li2020dlfix, li2022dear} formulates program repair as a translation task, learning to transform buggy code into correct patches using historical bug-fixes. Afterwards, leveraging larger-scale training corpora and advanced pre-training algorithms, PLM-based approaches \cite{xia2023plastic, wang2023rap, xia2023automated, xia2022less} demonstrate much more powerful bug-fixing performance.}

More recently, PLMs have expanded to Large Language Models (LLMs) with billions of parameters and training data, plus post-training with diverse reinforcement learning techniques to align with human preferences. As such, their utility has been examined in various software engineering tasks \cite{yang2024exploring, xue2024automated, xue2025classeval, xue2025new, yao2024survey}, including in the APR area \cite{xia2023automated, xia2024automated, yin2024thinkrepair, kong2025contrastrepair, hu2025tsapr, zhang2025repair}. 
{Specifically, according to their primary repair guidance mechanisms, LLM-based APR techniques can be broadly categorized into three paradigms:} (1) retrieval-based approaches (e.g., RepairAgent \cite{bouzenia2024repairagent} and ReinFix \cite{zhang2025repair}), which utilize external tools, e.g., static analysis and search mechanisms, to extract relevant repair ingredients and historical fixes for guiding patch generation; (2) feedback-based approaches (e.g., ChatRepair \cite{xia2024automated}, ContrastRepair \cite{kong2025contrastrepair}, and TSAPR \cite{hu2025tsapr}), which perform iterative self-correction to refine candidate patches with test feedback; and (3) hybrid approaches (e.g., ThinkRepair \cite{yin2024thinkrepair}), which synergistically combine the above two kinds of strategies for patch generation and refinement.


This article focuses on the second paradigm of LLM-based APR techniques, i.e., feedback-based approaches, and tries to resolve one of their significant limitations: the exclusive reliance on outcome-level failure symptoms, e.g., stack traces, for patch refinement.
Although these signals indicate where and how a failure is observed, they often fail to expose the intermediate runtime states that causally lead to the failure.
In practice, instead of merely relying on stack traces, developers typically observe program behavior by inserting instrumentation, e.g., print statements, thereby exposing intermediate runtime states and progressively localizing bug causes. As such, it is imperative to augment LLM-based APR tools with such a debugging-oriented cognitive process, thereby alleviating their reasoning burden when handling complex bugs. 


To address the above limitation, we propose DebugRepair, a self-directed debugging framework for LLM-based APR, which consists of three key components: 
\textbf{(1) Test Semantic Purification.} In practice, real-world unit tests often encapsulate multiple assertions targeting different testing scenarios of a focal method, not all of which are responsible for triggering the bug. To reduce useless debugging logs from executing irrelevant test scenarios in follow-up steps, we perform static program slicing on the failing test to extract the minimal semantic subset directly related to the failure. This purified test facilitates the follow-up instrumentation and debugging for LLMs.
\textbf{(2) Simulated Instrumentation.} When the LLM fails to repair a bug using only outcome-level failure symptoms, DebugRepair proactively guides it into a debugging phase. In this phase, the LLM autonomously identifies breakpoints and inserts instrumentation (e.g., print statements) along key execution paths according to the purified test. To ensure robust trace collection, we implement a hybrid strategy that supplements LLM with a deterministic, rule-based fallback. By executing the instrumented program, DebugRepair captures intermediate runtime states and transforms them into structured feedback that guides more precise patch refinement.
\textbf{(3) Debugging-Driven Conversational Repair.} Instead of isolated patch generation, we design a hierarchically iterative, conversation-driven repair framework to effectively utilize the collected runtime-state feedback. The outer loop is responsible for simulated instrumentation according to the purified test, while the inner loop assigns the LLM to refine candidate patches based on both prior fixing and newly acquired debugging information. The whole process continues until a plausible patch (i.e., one that passes all tests) is found or the predefined budget is exhausted.

{To evaluate the effectiveness of DebugRepair, we conduct extensive experiments on four distinct benchmarks: Defects4J (V1.2 and V2.0) \cite{just2014defects4j}, QuixBugs \cite{lin2017quixbugs}, HumanEval-Java \cite{jiang2023impact}, and RWB \cite{yin2024thinkrepair}.} Specifically, although DebugRepair follows a feedback-based paradigm of the LLM category, we include 15 baselines for comparison, covering template-based, learning-based, and LLM-based APR approaches. In particular, 6 of them are recently released SOTA LLM-based approaches across retrieval, feedback, and hybrid-based paradigms. Experimental results demonstrate that DebugRepair outperforms existing SOTA baselines. For example, with GPT-3.5 as the backbone model, DebugRepair correctly fixes 224 bugs on the Defects4J dataset, achieving an average improvement of 26.2\% over the SOTA LLM-based baselines.
Meanwhile, when evaluated with the DeepSeek-V3 backbone, DebugRepair continues to demonstrate its superiority by correctly fixing 295 bugs on the Defects4J dataset, achieving 59 more correct fixes than the second-best approach.
{Additionally, we implement DebugRepair on six LLMs from diverse families and sizes, showing that DebugRepair can fix 48.4\% more bugs than their vanilla settings on average, thereby demonstrating the model-agnostic nature of our proposal.}
Furthermore, the follow-up ablation study indicates that both the test purification and the simulated debugging mechanism play critical roles in enhancing LLMs’ ability to understand and repair complex bugs.
In summary, the main contributions of this paper are as follows:

\setlist[itemize]{leftmargin=1em, labelsep=0.6em, itemsep=2pt}

\begin{itemize}
    \item \textbf{Novel Technique.} We propose DebugRepair, a novel LLM-based APR framework that enhances patch generation via self-directed debugging. Unlike existing feedback-based APR approaches that rely mainly on outcome-level failure symptoms, DebugRepair introduces a debugging-oriented repair paradigm that equips LLMs with intermediate runtime evidence through test semantic purification, simulated instrumentation, and debugging-driven conversational repair.
    \item \textbf{Extensive Evaluation.} We conduct extensive experiments on four widely used APR benchmarks, namely Defects4J, QuixBugs, HumanEval-Java, and RWB, against 15 representative baselines spanning template-based, learning-based, and LLM-based APR techniques. The results show that DebugRepair consistently outperforms existing SOTA approaches across different backbone models. We further perform comprehensive ablation studies to validate the effectiveness of each component in the framework.
    \item \textbf{Open Science.} To facilitate reproducibility and future research, we release the implementation of DebugRepair and detailed repair results in \cite{146854hg69:online}.
\end{itemize}

\begin{figure*}[t]
    \centering
    {\includegraphics[width=1.0\linewidth]{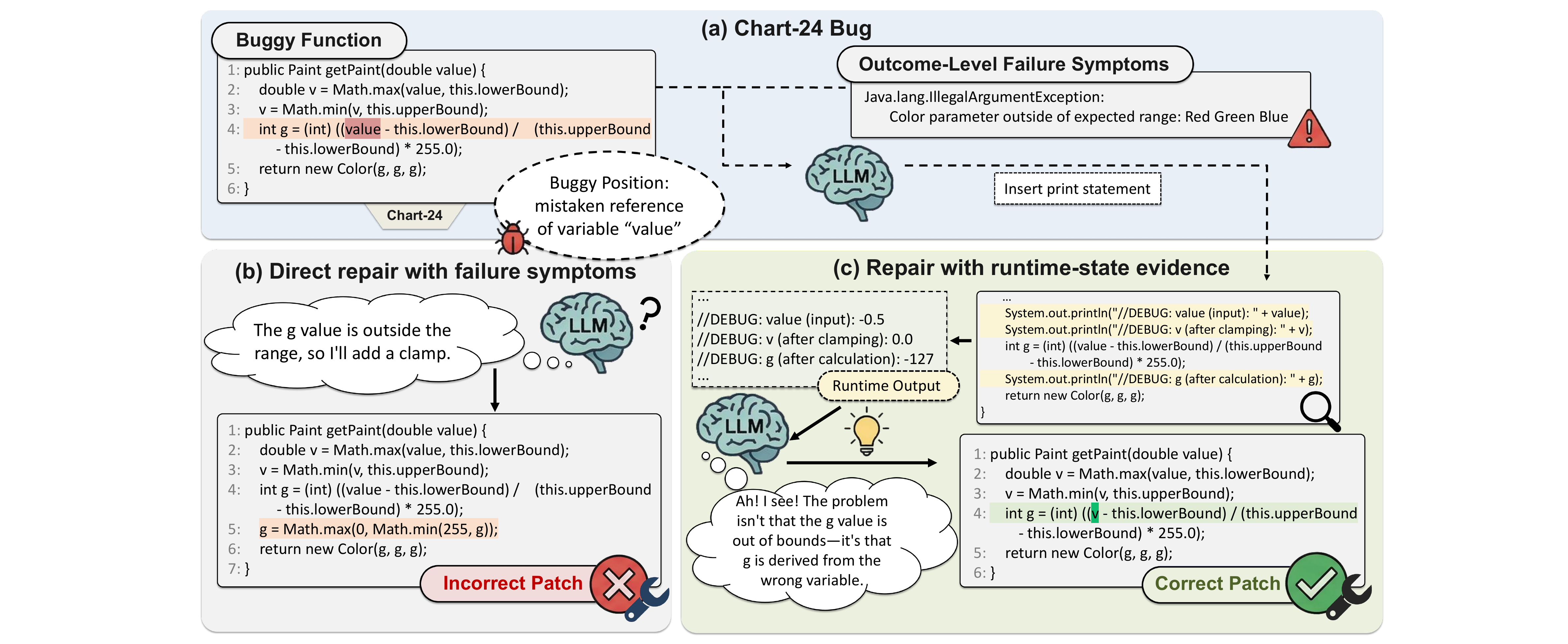}}
    \vspace{-0.3in}
    \caption{Motivation Example of DebugRepair.}
    \Description{}
    \label{fig:motivation}
    \vspace{-0.1in}
\end{figure*}
\section{Motivation}

In this section, we present a motivating example from Defects4J (Chart-24 bug) to illustrate the limitations of relying solely on outcome-level failure symptoms and the necessity of incorporating runtime-state evidence for effective program repair.

\begin{sloppypar}
\textbf{The Limitation of Outcome-Level Failure Symptoms.} Figure \ref{fig:motivation}(a) shows a buggy function \texttt{getPaint} designed to return an object \texttt{Color} according to a variable \texttt{v} clamped between \texttt{lowerBound} and \texttt{upperBound}, which is derived from the input parameter \texttt{value}.
However, the variable \texttt{g} for constructing \texttt{Color} object erroneously uses the raw input \texttt{value} instead of the restricted variable \texttt{v}, causing the program to crash with an exception, namely \texttt{java.lang.IllegalArgumentException}.
As shown in Figure \ref{fig:motivation}(b), existing LLM-based APR tools are typically fed buggy code and the above outcome-level failure symptoms to fix programs. However, such symptoms only confirm the manifestation of the failure (an invalid color parameter) while concealing the intermediate states, leading LLMs to conclude that \texttt{g} is outside the valid range and to incorrectly hypothesize that it simply needs to enforce a proper bound. Consequently, this leads to a plausible but incorrect patch, i.e., explicitly clamping \texttt{g} before sending to the constructor of \texttt{Color}, shown in Line 5. The incorrect patch merely masks the symptom rather than fixing the mistaken variable reference of \texttt{value} in Line 4.
\end{sloppypar}

\textbf{The Necessity of Debugging.} In contrast, consider how a human developer would fix this bug. As shown in Figure \ref{fig:motivation}(c), they would likely insert print statements to observe the runtime state, thereby visualizing the intermediate execution trace: \texttt{value} is -0.5, \texttt{v} is evaluated as 0.0, while \texttt{g} is calculated as -127.
By inspecting these runtime values, the root cause becomes immediately transparent for LLMs: while the variable \texttt{v} is correctly clamped to the legal lower bound (0.0), the subsequent calculation for \texttt{g} still yields an illegal negative value (-127). This conflict apparently reveals that \texttt{g} is computed from the raw, negative \texttt{value} (-0.5) rather than the properly restricted \texttt{v}, thereby helping the LLM generate the correct patch. This example demonstrates that runtime intermediate states act as a critical bridge between the symptom and the root cause. By enabling the LLM to proactively insert \texttt{print} statements and analyze the resulting debugging output, rather than passively consuming outcome-level failure symptoms, we can empower the model to precisely resolve the underlying logic errors.

\section{Approach}

\subsection{Framework Overview}

\FloatBarrier
\begin{figure*}[t]
    \centering
    \vspace{-0.1in}
    {\includegraphics[width=1.0\linewidth]{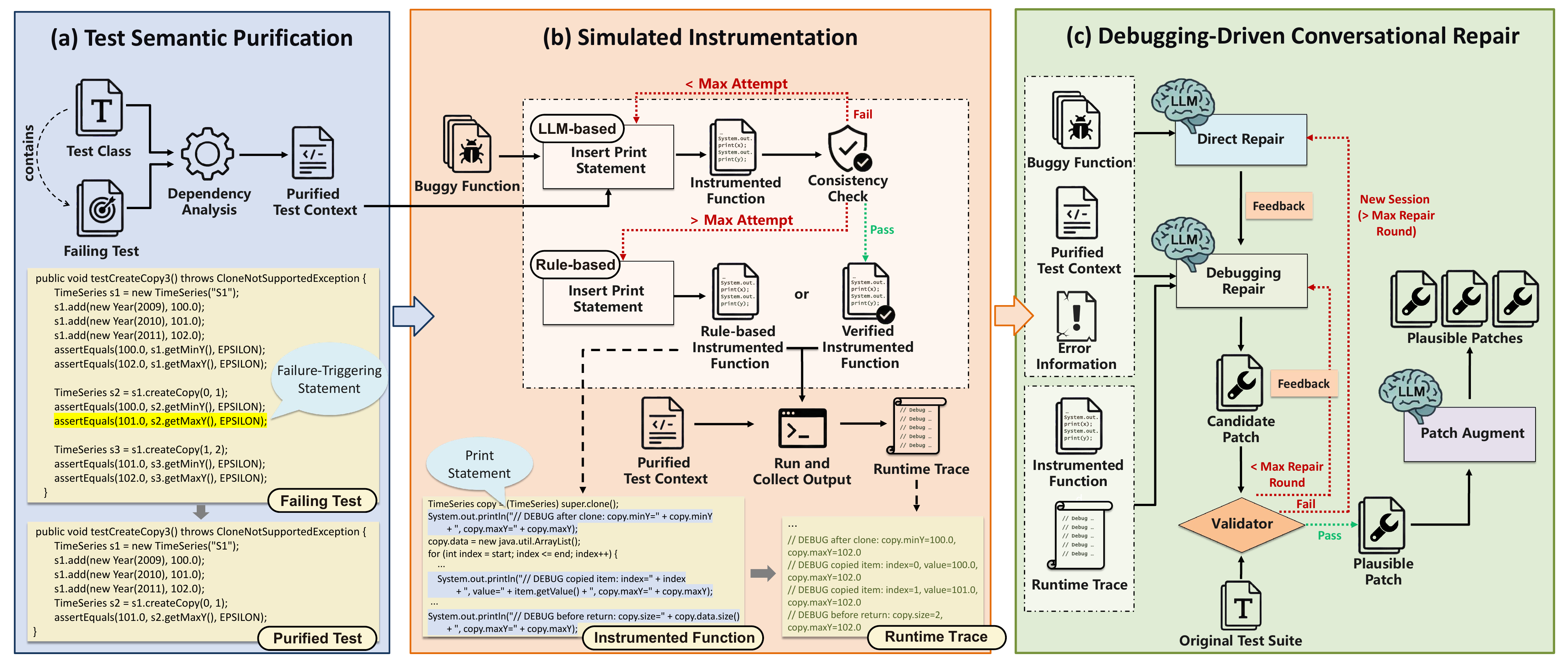}}
    \vspace{-0.3in}
    \caption{Overview of DebugRepair.}
    \Description{}
    \label{fig:framework}
    \vspace{-0.1in}
\end{figure*}

To bridge the gap between the LLM's understanding of static code and its perception of runtime behavior, we propose DebugRepair. This is an LLM-based APR framework designed to simulate the cognitive debugging process of human developers. As shown in Figure \ref{fig:framework}, the workflow of DebugRepair consists of three core phases:
\begin{itemize}[topsep=2pt,itemsep=1pt,parsep=0pt,partopsep=0pt]
    \item \textbf{Test Semantic Purification.}
    We perform semantic-level dependency analysis on the original failing test to strip away irrelevant test logic and noise, constructing a minimized test. This helps the LLM better focus on the current failure and reduce the irrelevant debugging logs in the follow-up phases.
    \item \textbf{Simulated Instrumentation.}
    {For complex bugs (i.e., those for which the current LLM's initial direct-repair attempt fails to produce a plausible patch), we proactively instrument the code using a hybrid strategy that combines LLM-based instrumentation with a rule-based fallback to inject print statements at critical code locations.} By executing the instrumented code within the purified test context, we capture the intermediate states of the program execution, thereby facilitating the program repair in follow-up phases.
    \item \textbf{Debugging-Driven Conversational Repair.} This phase proceeds in a hierarchical and iterative fashion, where an outer loop manages the simulated instrumentation and an inner loop performs the specific fixing based on the runtime traces generated above, thereby forming a debugging-driven multi-turn conversation to simulate human debugging practice.
\end{itemize}

\subsection{Test Semantic Purification}

Real-world test cases typically encapsulate multiple assertions and related dependencies to achieve comprehensive testing coverage. However, not every assertion triggers the current bug, yet retaining irrelevant ones may introduce noise into logs generated by instrumentation. To mitigate this interference for subsequent debugging, we propose a Test Semantic Purification approach, detailed in Algorithm \ref{algo:purification}.

Formally, given a failing test method $T$ parsed into a sequential list of statements $S = \langle s_1, s_2, \dots, s_n \rangle$ alongside the Abstract Syntax Tree (AST) of its enclosing test class $\mathcal{C}$ in which $T$ is defined, our objective is to extract a minimal statement subset $T_{min} \subset T$ along with its required external dependencies $\mathcal{D} \subset \mathcal{C}$. The algorithm begins by initializing a statement set \texttt{Slice} to compose $T_{min}$ and adding a failure-triggering statement $s_{fail}$ as its first element. Subsequently, we extract all variables and objects within $s_{fail}$ to form an identifier set, thereby initializing $V_{req}$ (Lines 1-3). Based on $V_{req}$, we trace backward anchored at $s_{fail}$ to search dependent statements from $S$, thereby expanding \texttt{Slice} (Lines 5-23).
Specifically, for each backward tracing, a preceding statement $s_i$ is inserted into the \texttt{Slice} if it satisfies either of the following two criteria (Lines 11-13): 

\begin{itemize}
    \item \textbf{Direct Data Dependency.}
    If variables or objects explicitly defined or assigned in $s_i$ (denoted as $V_{def}$) overlap with $V_{req}$ (i.e., $V_{def} \cap V_{req} \neq \emptyset$), $s_i$ is inserted in \texttt{Slice}.
    \item \textbf{Implicit State Modification.}
    In object-oriented languages such as Java, objects are frequently modified via method invocations rather than explicit assignment operators (e.g., \texttt{dataset.\allowbreak addValue(...)}). To capture these side effects, if $s_i$ is a non-assertion statement and the objects it utilizes (denoted as $V_{use}$) overlap with $V_{req}$ (i.e., $V_{use} \cap V_{req} \neq \emptyset$), $s_i$ is added in \texttt{Slice}.
\end{itemize}

\begin{figure*}[h]
    \centering
    \vspace{-0.1in}
    {\includegraphics[width=1.0\linewidth]{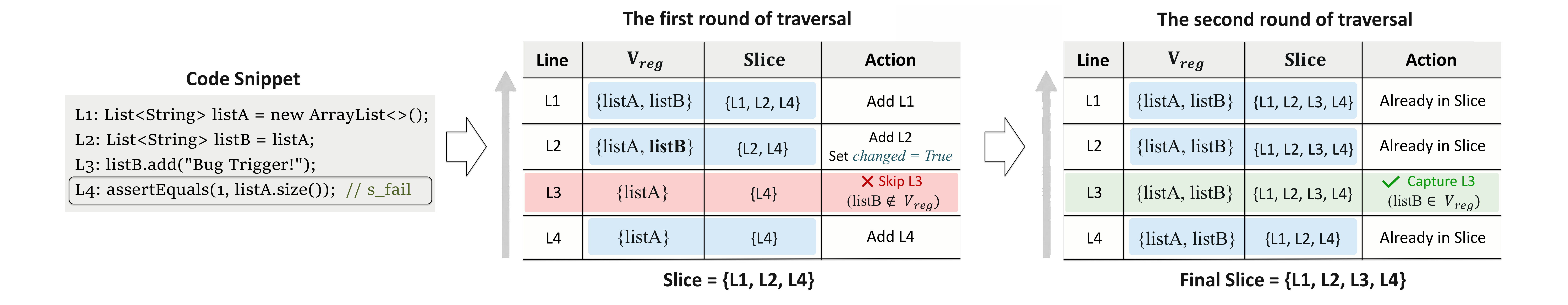}}
    \vspace{-0.3in}
    \caption{An example to illustrate the necessity of repeated traversal.}
    \Description{}
    \label{fig:slice}
    \vspace{-0.1in}
\end{figure*}

Whenever a statement $s_i$ is added to \texttt{Slice}, variables and objects that exist in it (either for definition or usage, denoted as $V_{i}$) are merged into $V_{req}$ to expand the searching scope of data dependencies (Lines 15 and 17).
Besides, since statements that are not dependent on $s_{fail}$ may also use objects in $V_{req}$, making the potentially caused side effects, in turn, affect $s_{fail}$. Thus, we have to repeat the traversal iteratively to update \texttt{Slice} and $V_{req}$ until no new objects appear, i.e., fixing the \texttt{changed} flag to \texttt{False} in the outer loop (Lines 18-19).
To illustrate the necessity of this iterative mechanism, Figure \ref{fig:slice} presents a motivating example, where \texttt{listB} is an alias of \texttt{listA} (L2). In this snippet, the assertion $s_{fail}$ at L4 fails because the operation \texttt{listB.add(...)} at L3 modifies their shared object, implicitly affecting \texttt{listA}.
When applying our slicing algorithm to this snippet, the first traversal among the statements preceding $s_{fail}$ (L4) will miss the implicit state modification at L3, because the object alias \texttt{listB} is not introduced into $V_{req}$ until L2 is processed.
However, the discovery of \texttt{listB} as a new object at L2 triggers the \texttt{changed} flag from \texttt{False} to \texttt{True} and induces a second traversal. As such, the algorithm rescans the snippet with the expanded $V_{req}$, successfully capturing the previously missed side effect at L3 and ultimately ensuring a complete \texttt{Slice}.


\newcommand{\algcomment}[1]{%
  \footnotesize\ttfamily\textcolor{gray}{#1}%
}
\SetCommentSty{algcomment}      

\begin{algorithm}[!t]
\small
\SetAlgoLined
\KwInput{$T$: the failing test method, $s_{fail}$: the failure-triggering statement, $\mathcal{C}$: the AST of the test class}
\KwOutput{$T_{min}$: the purified test method, $\mathcal{D}$: the set of external dependencies}
  
  $S := \langle s_1, s_2, \dots, s_n \rangle \leftarrow \texttt{ParseToStatements}(T)$; \tcp*{parse method into a statement sequence}
  $Slice := \{ s_{fail} \}$; \tcp*{initialize slice with the failing statement}
  $V_{req} := \texttt{FetchVarAndObj}(s_{fail})$; \tcp*{fetch required variables and objects}
  $changed := \texttt{True}$\;
   
  \While{$changed$}{
      $changed := \texttt{False}$\;
      \For{$i := \texttt{IndexOf}(s_{fail}, S) - 1$ \KwTo $1$}
      {\tcp*{traverse preceding statements backwards}
          \If{$s_i \in Slice$}{
              \textbf{continue}\;
          }
          $V_{def} := \texttt{FetchDefinedVarAndObj}(s_i)$; \tcp*{fetch variables or objects defined in $s_i$}
          $V_{use} := \texttt{FetchUsedObj}(s_i)$; \tcp*{fetch objects used in $s_i$}
           
          \If{$(V_{def} \cap V_{req} \neq \emptyset) \lor (\neg \texttt{IsAssert}(s_i) \land V_{use} \cap V_{req} \neq \emptyset)$}{
              $Slice := Slice \cup \{ s_i \}$\;
              $V_{i} := \texttt{FetchVarAndObj}(s_{i})$\;
              $V_{new} := V_{i} \setminus V_{req}$\;
              $V_{req} := V_{req} \cup V_{i}$\;
              
              \If{$\exists v \in V_{new} \text{ such that } \texttt{IsObject}(v)$}{
                  $changed := \texttt{True}$;
              }
          }
      }
  }
   
    $T_{min} := \text{Reconstruct}(T, \text{Sort}(Slice))$; \tcp*{Assemble statements in original execution order}
    
    $M_{\mathcal{C}} := \text{FetchMethodSigs}(\mathcal{C})$; \tcp*{Fetch signatures of all methods defined in $C$}
    $\mathcal{D}_{m} := \text{FetchCalleeSigs}(T_{min}) \cap M_{\mathcal{C}}$; \tcp*{Initial direct dependent methods}
    $M_{new} := \mathcal{D}_{m}$;
    
    \While{$M_{new} \neq \emptyset$}{
        $M_{new} := \left(\text{FetchCalleeSigs}(M_{new}) \cap M_{\mathcal{C}}\right) \setminus D_m$;
        \tcp*{Find newly discovered indirect dependencies}
        $\mathcal{D}_{m} := \mathcal{D}_{m} \cup M_{new}$\;
    }
    
    $V_{all} := \text{FetchVarAndObj}(T_{min}) \cup \text{FetchVarAndObj}(\mathcal{D}_{m})$; \tcp*{Extract all variables and objects}
    $\mathcal{D}_{v} := V_{all} \cap \text{FetchDeclaredFields}(\mathcal{C})$; \tcp*{Resolve class-level dependent fields}
    
    $\mathcal{D} := \text{FetchDefinitions}(\mathcal{D}_{m}, \mathcal{D}_{v}, \mathcal{C})$; \tcp*{Form the complete external dependencies}
    
    return $\langle T_{min}, \mathcal{D} \rangle$; \tcp*{return the purified context}

\caption{Test Semantic Purification}
\label{algo:purification}
\end{algorithm}

Subsequently, we reconstruct the minimal test method, $T_{min}$, by preserving the original method signature of $T$ and sequentially assembling the statements retained in \texttt{Slice} according to their original relative order (Line 24).
Furthermore, the reconstructed test method $T_{min}$ may still reference class-level fields or call helper methods defined in its enclosing test class $\mathcal{C}$. To construct a complete testing context without unnecessary noise, we further conduct the following steps: (1) fetching helper methods defined in $\mathcal{C}$ that are directly or indirectly dependent by $T_{min}$; (2) fetching class-level fields that are directly or indirectly dependent by $T_{min}$.
To realize step (1), we maintain a list, namely $\mathcal{D}_{m}$, containing signatures of dependent methods. The initial elements are methods invoked by $T_{min}$ while defined in $C$, namely the direct dependence of $T_{min}$ (Lines 25-26). Afterwards, we further iteratively fetch methods in $\mathcal{D}_{m}$ and extract their callees to make intersection with methods defined in $C$, thereby finding indirect dependencies of $T_{min}$. Signatures of found methods are added to $\mathcal{D}_{m}$, and this process continues until the intersection becomes empty (Lines 27-31). To accomplish step (2), we locate each method definition according to $\mathcal{D}_{m}$ and $T_{min}$, extracting their dependent variables and objects. Whereafter, we make an intersection with fields declared in $C$ to fetch directly or indirectly dependent fields of $T_{min}$, denoted as $\mathcal{D}_{v}$. Finally, we fetch those required field declarations and method definitions according to $\mathcal{D}_{m}$ and $\mathcal{D}_{v}$ from $C$, forming the complete external dependencies $D$.
By stripping away irrelevant contextual noise, this purified context ensures that the subsequent simulated instrumentation and repair are focused exclusively on the failure-triggering scenario, while reducing the length of irrelevant logs.



\subsection{Simulated Instrumentation}

Existing LLM-based APR approaches often rely on outcome-level failure symptoms (e.g., stack traces), lacking perception of the program's intermediate runtime states. DebugRepair overcomes this limitation by inserting instrumentation into original buggy code according to the above purified test. We formalize this process as follows:

\textbf{Breakpoint Prediction.}
We model breakpoint identification as a prediction task $\mathcal{T}_{v}$ that infers critical variables $\mathcal{V}_{crit}$ essential for deducing the root cause. Accordingly, the selection of breakpoints for instrumentation can be guided by $\mathcal{V}_{crit}$.
To be specific, the condition of $\mathcal{T}_{v}$ is a crash context, including the purified test $T_{min}$ alongside its required external dependencies $\mathcal{D}$, the buggy function $F_{buggy}$, and the fault location $L_{bug}$. This process is driven by an LLM $m$, such that $\mathcal{V}_{crit} = \mathcal{T}_{v}(T_{min}, \mathcal{D}, F_{buggy}, L_{bug}, m)$.

\begin{figure*}[h]
    \centering
    \vspace{-0.1in}
    {\includegraphics[width=1.0\linewidth]{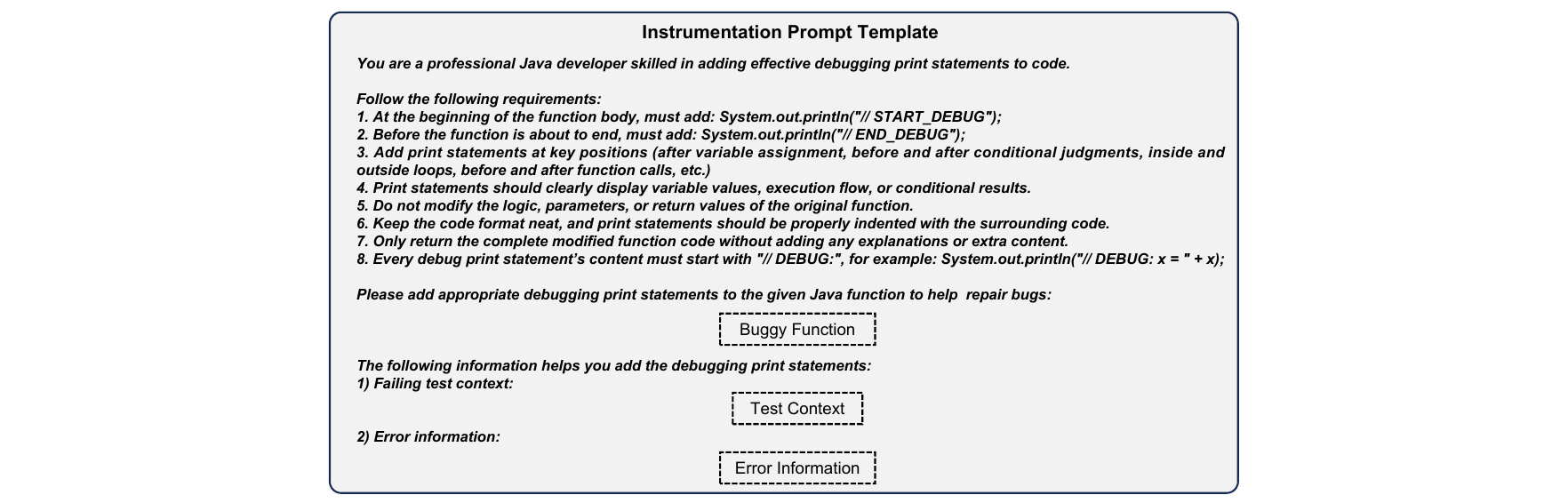}}
    \vspace{-0.3in}
    \caption{Instrumentation Prompt Template.}
    \Description{}
    \label{fig:instrumentation}
    \vspace{-0.1in}
\end{figure*}

\textbf{Semantics-Preserving Instrumentation.}
In this step, we primarily leverage the LLM $m$ to perform the instrumentation injection, denoted as $\mathcal{T}_{inst}$. Given the buggy function $F_{buggy}$ and the critical variable set $\mathcal{V}_{crit}$, $\mathcal{T}_{inst}$ prompts the LLM to generate an instrumented version $F_{inst}$ by inserting print statements (e.g., \texttt{System.out.println}) around $\mathcal{V}_{crit}$, i.e., $F_{inst}=\mathcal{T}_{inst}(F_{buggy}, \mathcal{V}_{crit}, m)$. The detailed insertion instructions are shown in Figure~\ref{fig:instrumentation}.

However, LLM-generated instrumentation may introduce unintended edits beyond print statements, such as auxiliary comments or even syntactic errors. To ensure that the instrumented code preserves the original program semantics, we impose a two-stage consistency check. First, we perform a line-wise equivalence check between normalized $F_{inst}$ and $F_{buggy}$, where the normalization process $\text{Norm}(\cdot)$ denotes a function that removes all print statements and comments from a given code. Hence, we can confirm that there is no additional statement modified or added to distort the code logic. 
Formally, we require:$\text{Norm}(F_{inst}) \equiv \text{Norm}(F_{buggy})$.
Second, we perform a compilation check on $F_{inst}$, because even if the above condition holds, the instrumentation may still break syntactic correctness.
Therefore, we compile $F_{inst}$ and accept it only if the compilation succeeds without errors. 

\begin{algorithm}[t]
\small
\SetAlgoLined
\KwInput{$F_{buggy}$: buggy function}
\KwOutput{$F_{inst}$: instrumented function}

$T := \texttt{ParseToAST}(F_{buggy})$\;
$M := \texttt{FetchMethodDecl}(T)$\;

\texttt{InsertPrintAfter}$(M,\ \texttt{"// START\_DEBUG"})$\;

\ForEach{$n \in \texttt{TraverseAST}(M)$}{
    \uIf{\texttt{IsVarInitStmt}$(n)$ or \texttt{IsVarAssigStmt}$(n)$}{
        $names := \texttt{FetchVarNames}(n)$\;
        $vals := \texttt{FetchVarVals}(names)$\;
        \texttt{InsertPrintAfterStmt}$(n,\ \texttt{"// DEBUG [VAR] "} + names.join(\texttt{","}) + \texttt{" = "} + vals.join(\texttt{","}))$\;
    }
    \uElseIf{\texttt{IsIfStmt}$(n)$}{
        $c := \texttt{FetchConditionExpr}(n)$\;
        $(s_1, s_2, t) := \texttt{HandleExpr}(c, \texttt{"// DEBUG [COND] "} + \texttt{ToString}(c) + \texttt{" = "})$\;
        \texttt{ReplaceConditionExpr}$(n, t)$\;
        \texttt{InsertStmtBeforeIf}$(n, [s_1; s_2])$\; 
    }
    \uElseIf{\texttt{IsWhileOrForStmt}$(n)$}{
        $c := \texttt{FetchConditionExpr}(n)$\;
        $(s_1, s_2, t) := \texttt{HandleExpr}(c, \texttt{"// DEBUG [LOOP] "} + \texttt{ToString}(c) + \texttt{" = "})$\;
        \texttt{ReplaceConditionExpr}$(n, \texttt{true})$; \tcp*{Replace the loop condition (c) with true.}
        $s_3 := \texttt{GenConBreakStmt}(\neg t)$;  \tcp{Generate an if-stmt, using non-t as the break condition.}     
        \texttt{InsertStmtAtLoopEntry}$(n, [s_1; s_2; s_3])$\;
    }
    \uElseIf{\texttt{IsReturnStmt}$(n)$}{
        $c := \texttt{FetchReturnExpr}(n)$\;
        $(s_1, s_2, t) := \texttt{HandleExpr}(c, \texttt{"// DEBUG [RETURN] "})$\;
        \texttt{ReplaceReturnExpr}$(n, t)$\;
        \texttt{InsertStmtBeforeReturn}$(n,\ [s_1; s_2])$\;
    }
    \uElseIf{\texttt{IsEmptyReturnStmt}$(n)$}{
        \texttt{InsertPrintBeforeReturn}$(n,\ \texttt{"// DEBUG [RETURN] void"})$\;
    }
}

$R := \texttt{FetchExitPos}(M)$\;
    \ForEach{$r \in R$}{
        \texttt{InsertPrintBeforeExit}$(r,\ \texttt{"// END\_DEBUG"})$\;
    }
$F_{inst} := \texttt{ASTToCode}(T)$\;
\Return{$F_{inst}$}\;
\SetKwFunction{MyFunction}{HandleExpr}   
\SetKwProg{Fn}{Function}{:}{\KwRet}        
\Fn{\MyFunction{$c, LOG$}}{                  
    $t := \texttt{CreateTempVar}()$\;
    $s_1 := \texttt{BuildTempAssignStmt}(t, c)$\;
    $s_2 := \texttt{GenPrintStmt}(LOG + t)$\;                 
    \KwRet{$s_1, s_2, t$}\;                          
}    
\caption{Rule-based Instrumentation}
\label{algo:rule-instrumentation}
\end{algorithm}

Once the LLM-instrumented code fails the above two-stage consistency check over a pre-defined maximum attempt limit ($M_{inst}$), a deterministic rule-based instrumentation strategy (detailed in Algorithm \ref{algo:rule-instrumentation}) will be activated as a fallback.
Specifically, given a buggy function $F_{buggy}$, we first parse it into an AST (Line 1) and locate the method declaration node to insert a print statement with ``// START\_DEBUG'', as the first statement of $F_{buggy}$ (Lines 2-3).
Next, we continue traversing the whole AST to insert instrumentation for other critical breakpoints and variables (Lines 4-27).
For each statement of variable initialization or assignment, we fetch all variables involved and insert a logging statement immediately after it to print the name and value of each variable being updated (Lines 5-8).
For each \texttt{if} statement, we extract the condition expression $c$ and insert a logging statement before it to record both the condition and its specific value (Lines 9-13). To avoid the unexpected changing of certain variables in $c$ owing to the repeated execution, we assign the value of $c$ to a temporal variable $t$ for logging, simultaneously replacing $c$ with $t$ as the condition expression of the \texttt{if} statement as well.
For \texttt{While} and \texttt{For} statements, we still have to resolve the repeated computation issue during the instrumentation, but we cannot simply replace the loop condition with a temporal variable as \texttt{If} statements, because it will be repeatedly computed to examine the reachability of loop boundary. Therefore, we first replace the loop condition with \texttt{True}, and then generate an assignment statement $s_1$ from the loop predicate $c$ to the temporal variable $t$. Afterwards, we generate a logging statement $s_2$ printing the loop condition expression and its associated value. In addition, we further generate a conditional break statement $s_3$ using non-$t$ for judgment, thereby controlling the termination inside the loop. All the above newly generated statements, namely $s_1$, $s_2$, and $s_3$, are inserted at the loop entry (Lines 14-19).
For non-empty return statements, we extract the return expression $c$ and assign its value to a temporal variable $t$ for rule-based logging. Likewise, $t$ is also used to substitute $c$ in the original return statement, thereby avoiding repeated computation (Lines 20-24).
For other empty return statements, i.e., return statements with no expression after them for function exit only, we insert a fixed logging statement before them to specify this purpose (Lines 25-26).
After all the above instrumentation is completed, we identify every explicit exit location and insert a fixed logging statement with ``// END\_DEBUG'' before them, thereby specifying the end of each function (Lines 28-31).
Finally, we deserialize the modified AST back into source code to obtain the instrumented function $F_{inst}$ (Lines 32-33). 
Although this instrumentation cannot uncover targeted state updates for different $F_{buggy}$, it still reveals critical breakpoints and variables that developers commonly inspect during debugging. Thus, it remains useful as a supplement to LLM-based instrumentation, especially when LLM-generated code is uncompilable. 


\textbf{Runtime Trace Capture.}
Executing the verified $F_{inst}$ within the context of $\langle T_{min}, \mathcal{D} \rangle$ produces a runtime trace:
\[
\tau_{runtime} = \langle log_1, \dots, log_k \rangle,
\]
where each log records a critical variable $v \in \mathcal{V}_{crit}$ and its value $val$ during execution before $s_{fail}$, thereby helping reveal data flow issues that cannot be identified from outcome-level failure symptoms alone. 

\subsection{Debugging-Driven Conversational Repair}

Leveraging the captured trace $\tau_{runtime}$, DebugRepair initiates a conversational repair process. Instead of simply repeated sampling or relying on the one-shot generation capability of LLMs, this phase adopts a debugging logic similar to that of human developers. Specifically, it employs a closed-loop mechanism of generation, verification, and feedback to allow the LLM to progressively approximate the correct patch over multi-turn conversations. The core advantage of this strategy is that it enables the model to utilize negative feedback from historical interactions to correct errors progressively, rather than starting the repair from scratch each time. To avoid getting stuck in local optima during patch generation, we design a hierarchical iterative mechanism consisting of debugging sessions and repair rounds.

\FloatBarrier
\begin{figure*}[t]
    \centering
    \vspace{-0.1in}
    {\includegraphics[width=1.0\linewidth]{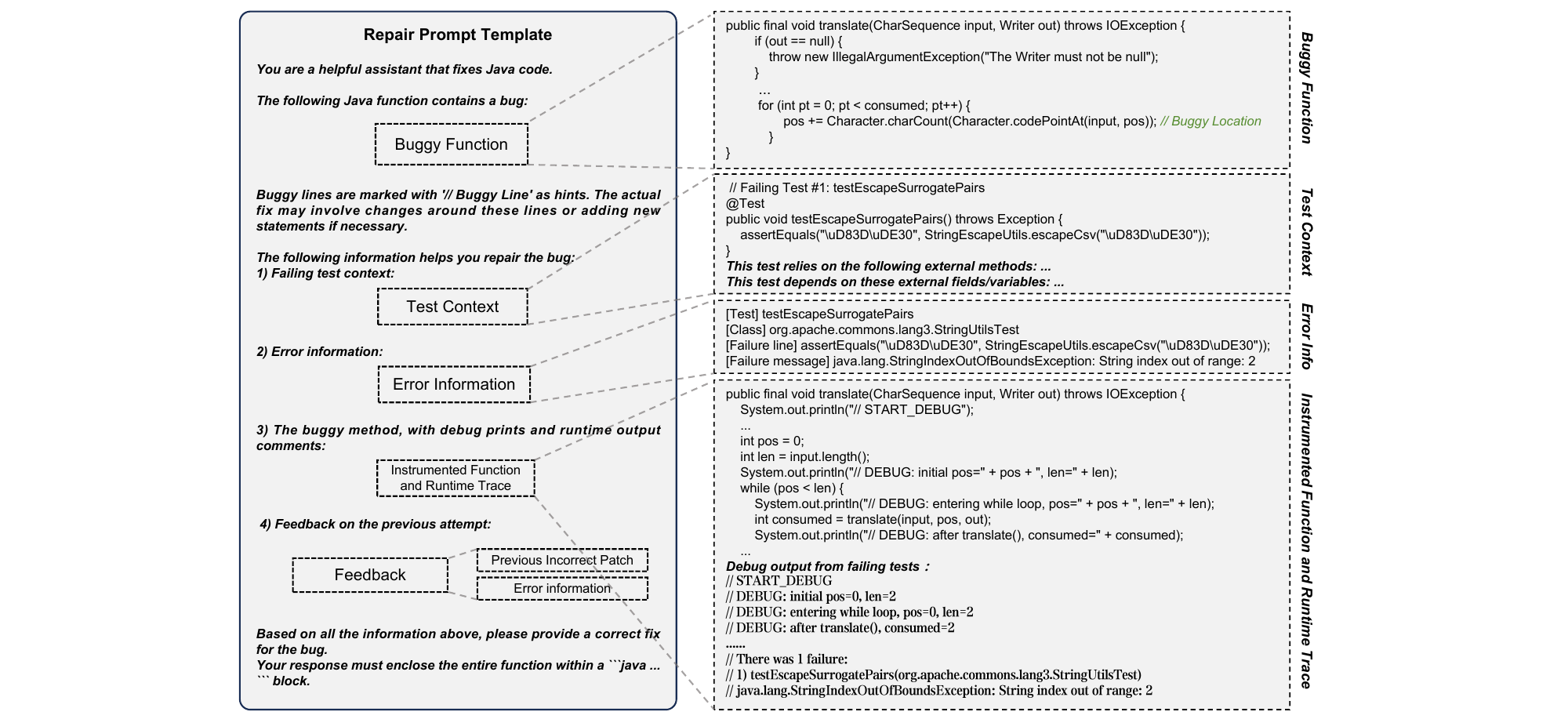}}
    \vspace{-0.3in}
    \caption{Illustration of the Prompt Construction.}
    \Description{}
    \label{fig:prompt}
    \vspace{-0.1in}
\end{figure*}

\textbf{Hierarchical Iteration Strategy.}
Let $N_{session}$ denote the maximum number of debugging sessions, and $K_{round}$ denote the maximum repair rounds within a single session. The process proceeds as follows:

\begin{itemize}
    \item \textbf{Session Initialization via Direct Repair.} At the beginning of each session, the LLM first attempts a direct repair using a basic prompt that contains the buggy function $F_{buggy}$ and the outcome-level failure symptoms. If this initial attempt yields a plausible patch, the session terminates. However, if the generated patch (denoted as $P_{0}$) fails the verification, DebugRepair captures the corresponding error message $E_{0}$ and transitions into the debugging mode for the subsequent rounds.
    \item \textbf{Debugging-Augmented Prompting.} Once the direct repair fails, we construct a comprehensive prompt for the debugging rounds. This augmented context incorporates the buggy function $F_{buggy}$, its instrumented version $F_{inst}$, the purified test context $\langle T_{min}, \mathcal{D} \rangle$, the serialized dynamic trace $\tau_{runtime}$, as well as the initial failed patch $P_{0}$ and its corresponding error message $E_{0}$. The specific repair prompt is shown in Figure \ref{fig:prompt}. This comprehensive context enables the LLM to perform evidence-based reasoning, deducing logic errors by contrasting observed runtime values against expected behaviors rather than relying solely on outcome-level failure symptoms.
    \item \textbf{Augmenting with Feedback History.} 
    In each subsequent round $k$ (where $1 \le k < K_{round}$), the LLM generates a candidate patch $P_{k}$ based on the above augmented context and the feedback history. If $P_{k}$ fails verification, the framework captures the specific failure symptoms, including the compilation or runtime errors.
    We then collect the conversation information $\langle P_{k}, E_{k} \rangle$ and append it to the feedback history. The LLM utilizes this updated history to generate a refined patch $P_{k+1}$ in the next round. This iterative correction continues until a plausible patch is found or the iteration limit $K_{round}$ is reached.
    \item \textbf{Session-Level Re-Debugging.} If a plausible patch is not found after $K_{round}$ attempts, it implies that the current debugging information might be insufficient or misleading. Consequently, the framework terminates the current session, discards the conversation history, and triggers a new debugging session to capture a fresh dynamic trace. 
    The whole process terminates when a plausible patch is found or the global budget ($N_{session} \times K_{round}$) is exhausted.
\end{itemize}

\begin{figure*}[h]
    \centering
    \vspace{-0.1in}
    {\includegraphics[width=1.0\linewidth]{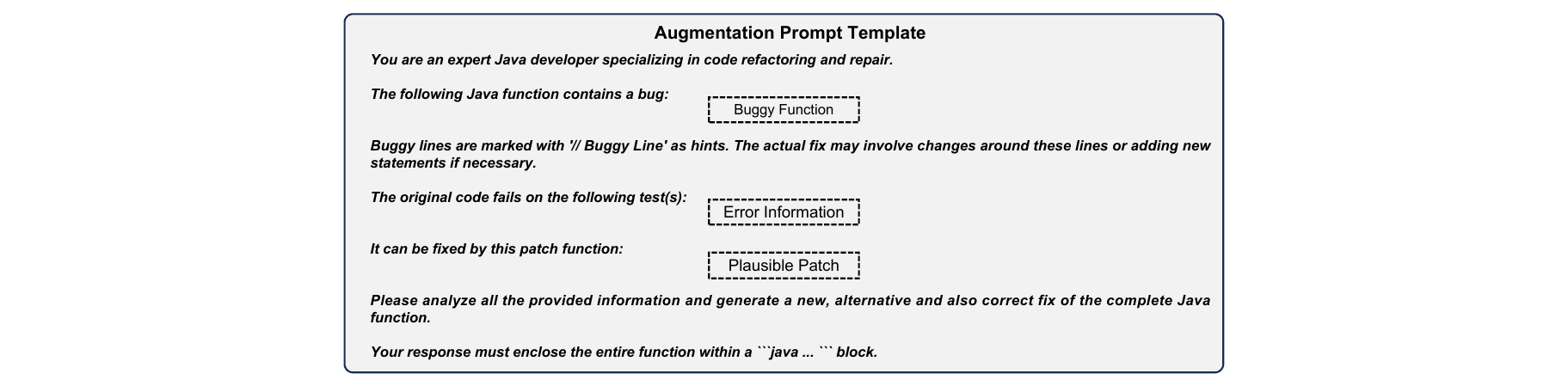}}
    \vspace{-0.3in}
    \caption{Patch Augmentation Prompt Template.}
    \Description{}
    \label{fig:prompt-aug}
    \vspace{-0.1in}
\end{figure*}

\textbf{Patch Augmentation.}
Finally, DebugRepair outputs a set of plausible patches.
While an initial plausible patch can successfully pass the entire test suite, it may not always represent the semantically correct fix. This discrepancy frequently arises from test overfitting, as incomplete test suites may fail to cover all intended program behaviors. To mitigate this risk, following prior APR work \cite{xia2024automated, kong2025contrastrepair}, we introduce a patch augmentation module. Instead of restarting the repair process from scratch, DebugRepair leverages the generated plausible patch as a valuable reference, given that both plausible and correct patches share the characteristic of satisfying the available test suite. {Specifically, as shown in Figure~\ref{fig:prompt-aug}, the augmentation prompt takes the buggy function, the corresponding error information, and the initial plausible patch as input. The LLM is then instructed to generate an alternative fix that follows a similar repair logic but is implemented differently.} All newly generated variants are subsequently validated against the test suite. Finally, DebugRepair outputs the resulting set of plausible patches.

\subsection{End-to-End Running Example}

\begin{figure*}[h]
    \centering
    \vspace{-0.1in}
    {\includegraphics[width=1.0\linewidth]{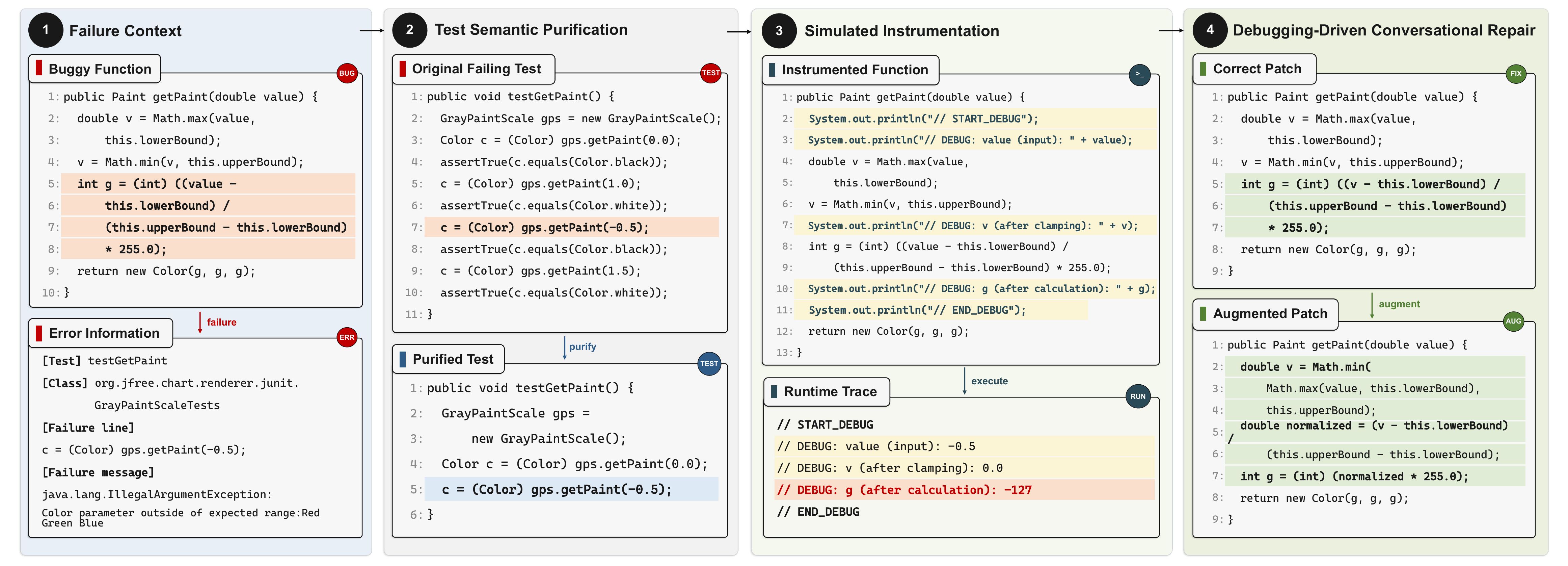}}
    \vspace{-0.3in}
    \caption{An end-to-end runtime example of DebugRepair on the Chart-24 bug.}
    \label{fig:running-example}
    \vspace{-0.1in}
\end{figure*}

{
Figure~\ref{fig:running-example} illustrates the complete execution of DebugRepair on the Chart-24 bug introduced in Figure~\ref{fig:motivation}. Given the buggy function and its failure information, Test Semantic Purification first removes test statements unrelated to the failure-triggering scenario while retaining the necessary context for invoking \texttt{getPaint(-0.5)}. Based on the purified test, Simulated Instrumentation inserts debugging statements around the key intermediate variables. Executing the instrumented function produces a runtime trace showing that the input \texttt{value} is $-0.5$, the clamped variable \texttt{v} becomes $0.0$, while \texttt{g} is still calculated as $-127$.

This runtime evidence reveals that \texttt{g} is incorrectly derived from the original \texttt{value} rather than the clamped variable \texttt{v}. Accordingly, Debugging-Driven Conversational Repair replaces \texttt{value} with \texttt{v} in the computation of \texttt{g}, producing the correct patch. After the patch passes the test suite, Patch Augmentation further generates an alternative implementation that preserves the same repair logic by explicitly introducing a normalized intermediate value. This example demonstrates how the three core phases of DebugRepair progressively transform failure information into concrete runtime evidence and ultimately guide the generation of correct and diverse patches.
}

\section{Experimental Setup}

\subsection{Research Questions}
\label{rqs}

We evaluate DebugRepair on the following research questions (RQs):
\begin{itemize}
    \item \textbf{RQ1: How does DebugRepair perform in comparison with SOTA APR  techniques?} 
    This question evaluates the overall repair effectiveness of DebugRepair and compares its performance against existing SOTA APR baselines across diverse datasets.
    \item \textbf{RQ2: How does DebugRepair perform across different repair scenarios?} 
    This question investigates DebugRepair’s performance in three distinct repair scenarios (i.e., single-function, single-hunk, and single-line bugs).
    \item \textbf{RQ3: To what extent does DebugRepair improve the repair effectiveness of vanilla LLMs for APR?} 
    This question investigates the generality of DebugRepair across different backbone LLMs of diverse families and sizes.
    \item \textbf{RQ4: What are the contributions of different components of DebugRepair in improving repair effectiveness?} This question investigates the individual contribution of key components of DebugRepair on the overall repair effectiveness.
    \item \textbf{RQ5: How well does DebugRepair generalize to bugs introduced after the LLM training data cutoff?} 
    {Considering the potential data leakage risk caused by LLMs, this question evaluates DebugRepair’s generalization ability on benchmarks constructed or released after the training data cutoff of the backbone model.}
\end{itemize}

\subsection{Benchmarks}
\label{bench}

To evaluate repair effectiveness, we conduct experiments on two widely adopted benchmarks: Defects4J \cite{just2014defects4j} and QuixBugs \cite{lin2017quixbugs}. Following prior APR studies \cite{xia2023automated, xue2024exploring, xia2024automated, yin2024thinkrepair, zhang2025repair, hu2025tsapr}, we consider two versions of Defects4J (V1.2 and V2.0) in the main experiments. Defects4J-V1.2 contains 395 real-world bugs, while Defects4J-V2.0 introduces an additional 440 new bugs. QuixBugs is a smaller yet popular benchmark, comprising 40 function-level bugs with both Java and Python versions. Consistent with existing work \cite{zhu2021syntax, xia2024automated, yin2024thinkrepair, hu2025tsapr, wu2026causalrepair}, we further categorize all bugs into three repair scenarios: Single-Function (SF), Single-Hunk (SH), and Single-Line (SL). Note that the single-line category is a subset of the single-hunk category, and the single-hunk category is a subset of the single-function category. In QuixBugs-Java, all single-hunk bugs correspond to single-line fixes, while in QuixBugs-Python, all fixes are single-line. Detailed statistics for each repair scenario are reported in Table \ref{tab:details_d4j}.


Additionally, to assess generalizability under reduced data leakage risk, we include the HumanEval-Java \cite{jiang2023impact} benchmark, which consists of 163 single-hunk bug cases. HumanEval-Java was released after the data collection period used to train GPT-3.5 \cite{ModelsOp28:online}, thereby reducing the potential risk of data leakage. In this benchmark, developers convert Python programs from HumanEval \cite{chen2021evaluating} together with their corresponding test cases into Java implementations and JUnit test cases, and then deliberately inject some bugs into these correct Java programs.
{
To further strengthen this evaluation, we also include RWB (Real-World Bugs) \cite{yin2024thinkrepair}, which consists of recent real-world bugs collected from high-quality open-source projects. RWB contains two versions: RWB-V1.0 consists of 44 SF bugs collected from bug-fixing commits after October 2021, while RWB-V2.0 consists of 29 SF bugs collected after March 2023. Since the pre-training data of GPT-3.5 was collected before September 2021 \cite{ModelsOp28:online}, both RWB versions provide a temporally separated setting for evaluating DebugRepair with reduced data leakage risk.
}


\begin{table}[h]
\scriptsize
\vspace{-0.1in}
\caption{Statistics of Studied Benchmarks.}
\vspace{-0.1in}
\label{tab:details_d4j} 
\begin{tabular}{@{}lcccc@{}}
\toprule
Benchmarks      & \# Total Bugs & \# SF Bugs & \# SH Bugs & \# SL Bugs \\ \midrule
Defects4J-V1.2  & 391           & 255        & 154        & 80         \\
Defects4J-V2.0  & 444           & 228        & 159        & 78         \\
QuixBugs-Java   & 40            & 40         & 37         & 37         \\
QuixBugs-Python & 40            & 40         & 40         & 40         \\
HumanEval-Java  & 163           & 163        & 163        & 74         \\
RWB-V1.0        & 44            & 44         & 25         & 9          \\
RWB-V2.0        & 29            & 29         & 18         & 6          \\ \bottomrule
\end{tabular}
\vspace{-0.2in}
\end{table}

\subsection{Baselines}

To make a comprehensive evaluation on Defects4J and QuixBugs, we compare DebugRepair against 15 SOTA baselines across different categories, including eight learning-based approaches (CURE \cite{jiang2021cure}, Recoder \cite{zhu2021syntax}, SelfAPR \cite{ye2022selfapr}, RewardRepair \cite{ye2022neural}, KNOD \cite{jiang2023knod}, AlphaRepair \cite{xia2022less}, FitRepair \cite{xia2023plastic}, and RAP-Gen \cite{wang2023rap}), and one template-based traditional method (TBar \cite{liu2019tbar}). Furthermore, we include six LLM-based approaches, which are categorized into three subgroups: feedback-based approaches (ChatRepair \cite{xia2024automated}, ContrastRepair \cite{kong2025contrastrepair}, and TSAPR \cite{hu2025tsapr}), retrieval-based approaches (RepairAgent \cite{bouzenia2024repairagent} and ReinFix \cite{zhang2025repair}), and hybrid approaches (ThinkRepair \cite{yin2024thinkrepair}).
Additionally, following prior work \cite{xia2024automated, kong2025contrastrepair, yin2024thinkrepair}, we create an LLM-based baseline named BaseChatGPT, which directly uses a basic prompt (see details in our repository \cite{146854hg69:online}) without providing additional feedback, serving as a foundational comparison. 
{For comparisons using GPT-3.5, following the common practice in the APR community \cite{xia2023plastic, xia2022less, xia2024automated, zhu2021syntax, kong2025contrastrepair, yin2024thinkrepair}, we report the results provided by their original papers. Additionally, to make comparisons on the latest LLMs, we manually reproduce the  above six LLM-based baselines using DeepSeek-V3.}

\subsection{Evaluation Metrics}

Following previous work~\cite{jiang2023knod, jiang2021cure, kong2025contrastrepair, yin2024thinkrepair, zhang2025repair, hu2025tsapr}, we consider two widely used metrics to evaluate the effectiveness of both DebugRepair and baselines:

\begin{itemize}
    \item \textbf{Number of Plausible Fixes (\# Plausible):}
    Counts the number of bugs which can pass all the test cases after fixing, without further verification.
    \item \textbf{Number of Correct Fixes (\# Correct):}
    {Counts the number of bugs for which at least one generated plausible patch is judged to be semantically correct. Specifically, two researchers independently review each plausible patch to determine whether it correctly addresses the bug and is semantically consistent with the intended behavior represented by the developer patch. When disagreements arise, a third independent researcher conducts an additional review and moderates the discussion until consensus is reached.}
\end{itemize}

\subsection{Implementation}

For our experiments, we primarily adopt gpt-3.5-turbo (referred to as GPT-3.5) \cite{ModelsOp28:online} as the backbone model for DebugRepair, using the API provided by OpenAI. 
{Additionally, we further incorporate five other LLMs, including GPT-5.5, DeepSeek-V3, Qwen2.5-7B, Qwen2.5-Coder-7B, and Qwen2.5-32B. GPT-5.5 is accessed through the official OpenAI API, while the remaining four models are accessed through APIs provided by SiliconFlow \cite{SiliconF25:online}.} 
These models cover a wide range of architectures and parameter sizes, enabling a comprehensive evaluation across different LLM backbones.
Following prior work \cite{xia2024automated, kong2025contrastrepair, yin2024thinkrepair, zhang2025repair, wu2026causalrepair}, a sampling temperature of 1.0 is utilized to obtain a diverse set of potential patches. For fault localization (FL), to avoid potential biases introduced by FL tools, we align with recent studies \cite{xia2024automated, kong2025contrastrepair, yin2024thinkrepair, hu2025tsapr, zhang2025repair, wu2026causalrepair} and adopt the perfect fault localization setting.
Regarding the repair budget, we configure the number of debugging sessions ($N_{session}$) to 6, with a maximum of 4 repair rounds ($K_{round}$) per session. We further allow 8 additional queries in the patch augmentation stage. As a result, the maximum number of candidate patches explored for each bug is bounded by 32 (i.e., patch size = $6 \times 4 + 8$). Additionally, we set the maximum attempt threshold for the LLM instrumentation step ($M_{inst}$) to 10.
All experimental evaluations are conducted on a server running Ubuntu 20.04 with two Intel Xeon Gold 6138 CPUs and 251 GB of RAM. AST-based parsing and code manipulation for both Java and Python programs are implemented using the \textit{tree-sitter} \cite{treesitt90:online}. 

\section{Experimental Results}

In this section, we present the experimental results to answer the research questions formulated in Section \ref{rqs}. {We systematically evaluate the effectiveness, robustness, and generalizability of DebugRepair on widely used benchmarks (Defects4J, QuixBugs, HumanEval-Java, and RWB).}

\subsection{RQ1: Comparison with SOTA Approaches}

To answer RQ1, we evaluate the overall repair effectiveness of DebugRepair by comparing it against 15 SOTA baselines on the Defects4J and QuixBugs benchmarks. Table \ref{tab:result_overall} presents the comparative results with GPT-3.5 in terms of the number of plausible and correct patches.

\begin{table}[h]
\scriptsize
\setlength{\tabcolsep}{7pt}
\caption{Comparison of DebugRepair and baselines on Defects4J and QuixBugs (\# Correct/\# Plausible).}
\label{tab:result_overall} 
\begin{tabular}{@{}ccccccccc@{}}
\toprule
\multicolumn{2}{c}{\multirow{2}{*}{Category}}                      & \multirow{2}{*}{APR Approach}                                   & \multicolumn{1}{l}{\multirow{2}{*}{Patch Size}} & \multicolumn{3}{c}{Defects4J} & \multicolumn{2}{c}{QuixBugs} \\ \cmidrule(l){5-9} 
\multicolumn{2}{c}{}                                               &                                                               & \multicolumn{1}{l}{}                            & V1.2     & V2.0     & Total   & Java         & Python        \\ \midrule
\multicolumn{2}{c}{Template-based}                                 & TBar \cite{liu2019tbar}                      & -                                               & 68/95    & 8/25     & 76/120  & -            & -             \\ \midrule
\multirow{8}{*}{Learning-based} & \multirow{5}{*}{NMT-based}       & CURE \cite{jiang2021cure}                    & 5000                                            & 57/-     & 19/-     & 76/-    & 26           & -             \\
                                &                                  & Recoder \cite{zhu2021syntax}                 & 100                                             & 71/-     & 19/46    & 90/-    & 31           & -             \\
                                &                                  & SelfAPR \cite{ye2022selfapr}                 & 150                                             & 65/74    & 45/47    & 110/121 & -            & -             \\
                                &                                  & KNOD \cite{jiang2023knod}                    & 1000                                            & 71/85    & 50/85    & 121/170 & 25           & -             \\
                                &                                  & RewardRepair \cite{ye2022neural}             & 200                                             & 45/-     & 45/-     & 90/-    & 20           & -             \\ \cmidrule(l){2-9} 
                                & \multirow{3}{*}{PLM-based}       & AlphaRepair \cite{xia2022less}               & 5000                                            & 74/109   & 36/-     & 110/-   & 28           & 27            \\
                                &                                  & FitRepair \cite{xia2023plastic}                                                     & 4000                                            & 89/-     & 44/-     & 133/-   & -            & -             \\
                                &                                  & RAP-Gen  \cite{wang2023rap}                                                      & -                                               & 72/-     & 53/-     & 125/-   & -            & -             \\ \midrule
\multirow{8}{*}{LLM-based}      & Basic                            & BaseChatGPT                                                   & 32                                              & 75/104   & 67/100   & 142/204 & 33           & 32            \\ \cmidrule(l){2-9} 
                                & \multirow{2}{*}{Retrieval-based} & RepairAgent \cite{bouzenia2024repairagent}   & 117                                             & 92/96    & 72/90    & 164/186 & -            & -             \\
                                &                                  & ReinFix \cite{zhang2025repair}               & 45                                              & 104/-    & 109/-    & 213/-   & -            & -             \\ \cmidrule(l){2-9} 
                                & Hybrid                           & ThinkRepair \cite{yin2024thinkrepair}        & 125                                             & 98/-     & 107/-    & 205/-   & 39           & \textbf{40}            \\ \cmidrule(l){2-9} 
                                & \multirow{4}{*}{Feedback-based}  & ChatRepair \cite{xia2024automated}           & 500                                             & {\textbf{114}/-}    & 48/-     & 162/-   & 39           & \textbf{40}            \\
                                &                                  & ContrastRepair \cite{kong2025contrastrepair} & 160                                             & 103/-    & 40/-     & 143/201 & \textbf{40}           & \textbf{40}            \\
                                &                                  & TSAPR \cite{hu2025tsapr}                   & 32                                              & {108/\textbf{146}}  & 93/134   & 201/280 & \textbf{40}           & -             \\
                                &                                  & DebugRepair                                                   & 32                                              & 111/\textbf{146}  & \textbf{113}/\textbf{137}  & \textbf{224}/\textbf{283} & \textbf{40}           & \textbf{40}            \\ \bottomrule
\end{tabular}
\begin{tablenotes}
\footnotesize
\item $^{\Phi}$ ``-'' indicates no results reported in the original work. In addition, the highest numbers of plausible and correct fixes are highlighted in \textbf{bold}.
\end{tablenotes}
\end{table}


\textbf{Results on Defects4J.} 
As presented in Table \ref{tab:result_overall}, DebugRepair demonstrates superior repair capabilities, correctly fixing a total of 224 bugs on the Defects4J dataset, where versions 1.2 and 2.0 account for 111 and 113 fixes, respectively. Compared with ReinFix, the second-ranked approach, DebugRepair successfully fixes 11 additional bugs with a smaller patch size for patch generation.
In the realm of feedback-based APR, DebugRepair exhibits substantial performance gains with fewer or equal patch sizes. It correctly fixes 62 and 81 more bugs than ChatRepair and ContrastRepair, representing improvements of 38.3\% and 56.6\%, respectively. Furthermore, it fixes 23 more bugs than TSAPR, an 11.44\% increase over that tool's strong baseline of 201 fixes. Furthermore, our approach substantially outperforms learning-based approaches, achieving 68.4\%--194.7\% more correct fixes while requiring only 1/3 to 1/156 of the patch sizes.
Regarding bug fixes across different projects (as shown in Table \ref{tab:result_project}), DebugRepair achieves the highest performance on 7 out of 17 projects. In particular, DebugRepair achieves the sole best performance on five projects, including 37 correct fixes in Math, 29 in JacksonDatabind, 27 in Lang, 18 in Compress, and 6 in Gson, significantly outperforming all other tools on these projects. These results further demonstrate the effectiveness of DebugRepair in APR.
{Additionally, to enable comparisons using one of the latest and more capable LLMs, we manually reproduced the six LLM-based baselines (i.e., ChatRepair, ContrastRepair, ThinkRepair, RepairAgent, TSAPR, and ReinFix) on Defects4J using DeepSeek-V3 as the backbone model. As shown in Table~\ref{tab:result_deepseek}, DebugRepair correctly fixes 295 bugs, outperforming the strongest baseline, ReinFix (236 fixes), by 59 correct fixes (25.0\%). It also achieves the highest number of correct fixes on both Defects4J-V1.2 and V2.0, with 139 and 156 fixes, respectively. These results further demonstrate that the advantage of DebugRepair persists when evaluated with a newer and stronger backbone model.}

\begin{table}[h]
\scriptsize
\caption{Number of correct fixes across different projects on Defects4J.}
\label{tab:result_project} 
\resizebox{1.0\columnwidth}{!}{
\begin{tabular}{@{}ccccccccccccccccccc@{}}
\toprule
Approach                         & Chart       & Closure     & Lang        & Math        & Mockito    & Time       & Cli         & Codec       & Collect    & Compress    & Csv        & Gson       & Core       & Databind    & Xml        & Jsoup       & JxPath     & Total        \\ \midrule
\# Bugs                         & 26          & 174         & 64          & 106         & 38         & 26         & 39          & 18          & 4          & 47          & 16         & 18         & 26         & 112         & 6          & 93          & 22         & 835          \\ \midrule
RepairAgent                     & 11          & 27          & 17          & 29          & 6          & 2          & 8           & 9           & \textbf{1} & 10          & 6          & 3          & 5          & 11          & 1          & 18          & 0          & 164          \\
ThinkRepair                     & 11          & 34          & 19          & 27          & 6          & \textbf{4} & 9           & \textbf{10} & 0          & 16          & \textbf{8} & 5          & \textbf{7} & 17          & \textbf{2} & \textbf{28} & \textbf{2} & 205          \\
ChatRepair                      & \textbf{15} & \textbf{37} & 21          & 32          & 6          & 3          & 5           & 8           & 0          & 2           & 3          & 3          & 3          & 9           & 1          & 14          & 0          & 162          \\
ContrastRepair                  & 12          & 32          & 19          & 30          & \textbf{8} & 2          & 4           & 5           & 0          & 2           & 3          & 1          & 3          & 7           & 1          & 14          & 0          & 143          \\
TSAPR                           & 12         & 28          & 24          & 32          & \textbf{8} & \textbf{4} & \textbf{12} & 5           & 0          & 15          & 7          & 4          & 4          & 18          & 1          & 26          & 1          & 201          \\ \midrule
\multicolumn{1}{l}{DebugRepair} & 12          & 25          & \textbf{27} & \textbf{37} & 7 & \textbf{4} & 10           & 8           & 0          & \textbf{18} & 7          & \textbf{6} & 4          & \textbf{29} & \textbf{2} & 27          & 1 & \textbf{224} \\ \bottomrule
\end{tabular}
}
\begin{tablenotes}
\footnotesize
\item $^{\Phi}$ Note that ReinFix is excluded from this specific comparison because its GPT-3.5 performance under the single-function (SF) repair scenario across specific projects is neither reported nor open-sourced. In addition, the highest numbers of correct fixes are highlighted in \textbf{bold}.
Besides, Core is short for JacksonCore, Xml is short for JacksonXml, Databind is short for JacksonDatabind, and Collect is short for Collections. 
\end{tablenotes}
\vspace{-0.1in}
\end{table}

\begin{table}[h]
\scriptsize
\setlength{\tabcolsep}{7pt}
\caption{Repair results of DebugRepair and LLM-based baselines on Defects4J using DeepSeek-V3 (\# Correct/\# Plausible).}
\label{tab:result_deepseek}
\centering
\begin{threeparttable}
\begin{tabular*}{\columnwidth}{@{\extracolsep{\fill}}ccccccccc@{}}
\toprule
\multirow{2}{*}{Category} 
& \multirow{2}{*}{APR Approach} 
& \multicolumn{3}{c}{Defects4J-V1.2} 
& \multicolumn{3}{c}{Defects4J-V2.0} 
& \multirow{2}{*}{Total} \\ 
\cmidrule(lr){3-5} \cmidrule(lr){6-8}
& 
& SF 
& SH 
& SL 
& SF 
& SH 
& SL 
& \\ 
\midrule

Basic 
& BaseDeepSeek 
& 82/112 
& 59/76 
& 36/46 
& 73/91 
& 56/70 
& 31/40 
& 155/203 \\ 
\midrule

\multirow{2}{*}{Retrieval-based} 
& RepairAgent 
& 85/93 
& 67/73 
& 48/52 
& 87/94 
& 72/78 
& 45/48 
& 172/187 \\

& ReinFix 
& 118/\underline{181} 
& 83/120 
& 47/\underline{66} 
& \underline{118}/\underline{171} 
& \underline{89}/\underline{128} 
& 47/63 
& \underline{236}/\underline{352} \\ 
\midrule

Hybrid 
& ThinkRepair 
& 87/122 
& 66/89 
& 38/48 
& 93/129 
& 68/98 
& 35/51 
& 180/251 \\ 
\midrule

\multirow{4}{*}{Feedback-based} 
& ChatRepair 
& 103/156 
& 51/67 
& 45/56 
& - 
& - 
& 42/53 
& 145/209 \\

& ContrastRepair 
& \underline{133}/\textbf{182} 
& 86/\textbf{124} 
& \textbf{60}/\textbf{69} 
& - 
& - 
& \underline{48}/\textbf{66} 
& 181/248 \\

& TSAPR 
& 108/146 
& \underline{87}/115 
& 50/\underline{66} 
& 116/150 
& \underline{89}/114 
& 46/63 
& 224/296 \\

& DebugRepair 
& \textbf{139}/179 
& \textbf{98}/\underline{122} 
& \underline{57}/65 
& \textbf{156}/\textbf{180} 
& \textbf{113}/\textbf{129} 
& \textbf{61}/\underline{65} 
& \textbf{295}/\textbf{359} \\ 
\bottomrule

\end{tabular*}
\begin{tablenotes}
\footnotesize
\item $^{\Phi}$ ``-'' denotes unavailable SF/SH results for ChatRepair and ContrastRepair on Defects4J-V2.0, since their original studies evaluate only the SL scenario; our reproduction follows the same setting. In addition, the highest numbers of plausible and correct fixes are highlighted in \textbf{bold}, while the second-ranked ones are highlighted in \underline{underlines}.
\end{tablenotes}
\end{threeparttable}
\end{table}

\textbf{Results on QuixBugs.} 
We further evaluate DebugRepair on the QuixBugs benchmark. The results show that DebugRepair is capable of correctly fixing all 40 bugs in both the Java and Python datasets. 
{This exceptional performance matches the best results achieved by existing SOTA baselines, indicating its effectiveness across different datasets and providing further evidence that DebugRepair can be effectively adapted to Python programs.}

\begin{figure}[htbp]
\centering
    \subfigure[Venn on Defects4J-V1.2]{
        \includegraphics[width=0.45\columnwidth]{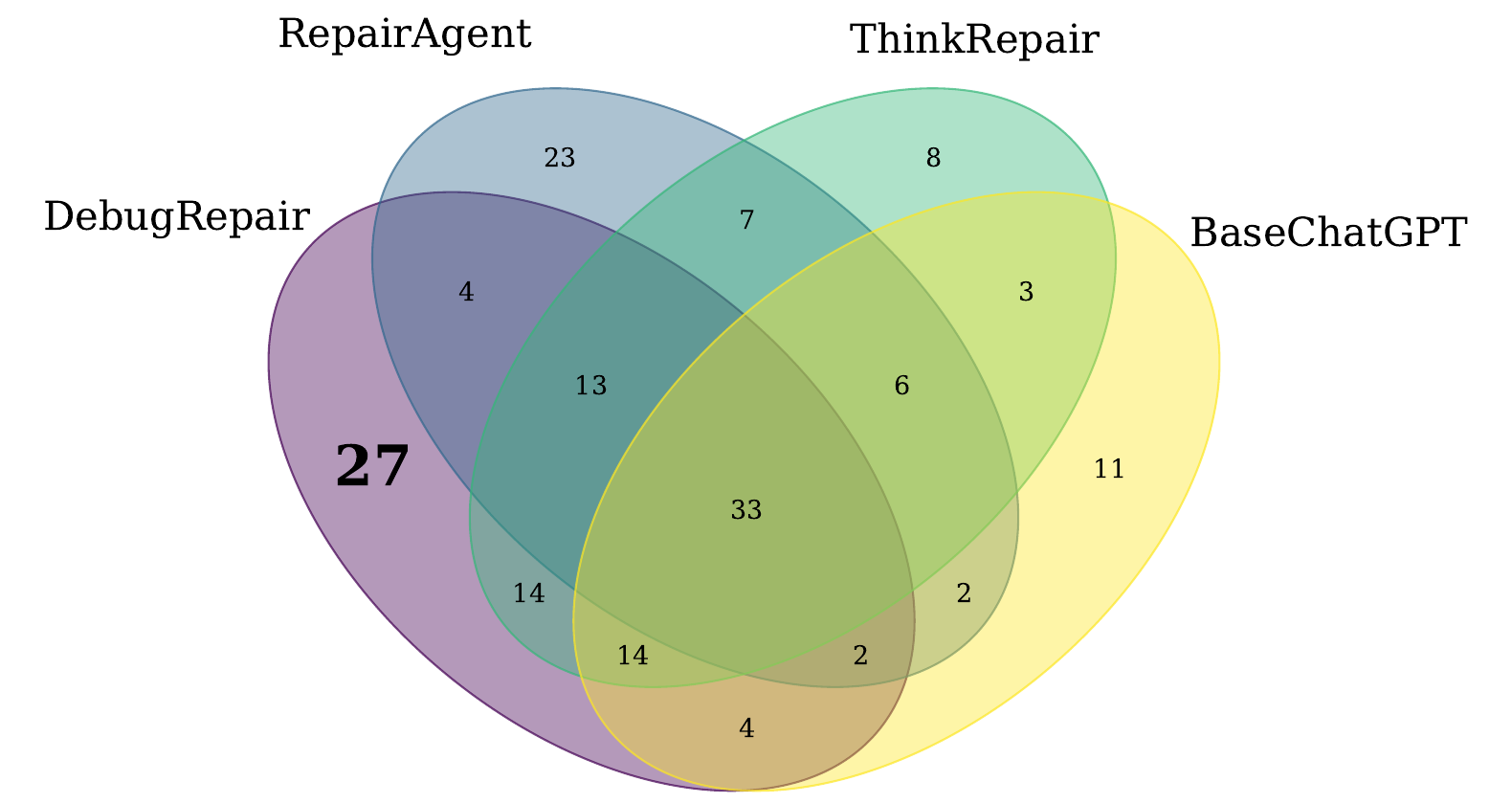}
        \label{venn_d4j12_1}
    }
    \subfigure[Venn on Defects4J-V2.0] {
        \includegraphics[width=0.45\columnwidth]{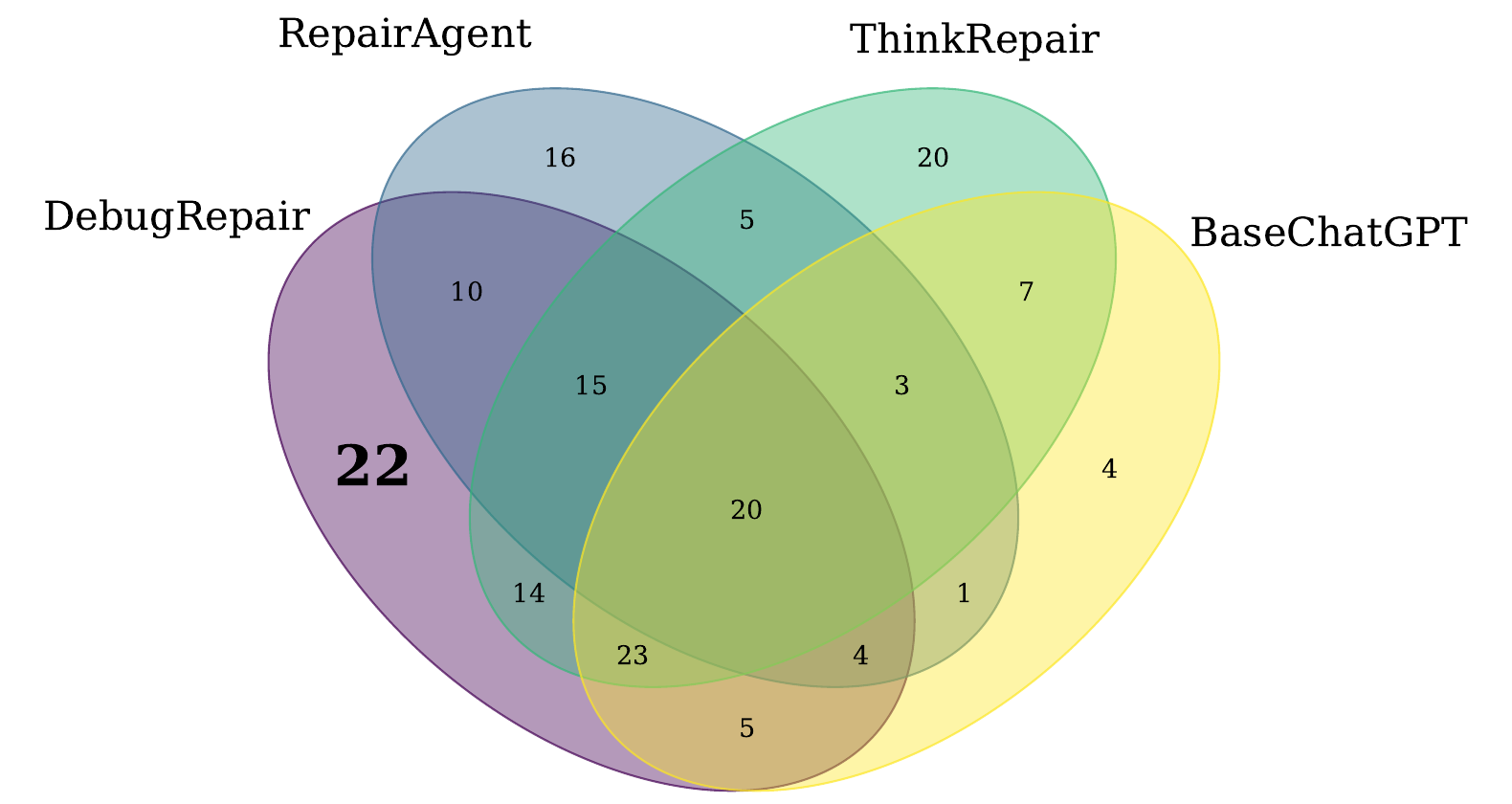}
        \label{venn_d4j20_1}
    }
    \vspace{-0.15in}
    \caption{Bug Fix Venn Diagram on Defects4J (DebugRepair, RepairAgent, ThinkRepair, BaseChatGPT)}
    \Description{}
    \label{d4j_venn_1}
\end{figure}

\begin{figure}[htbp]
\centering
    \subfigure[Venn on Defects4J-V1.2]{
        \includegraphics[width=0.4\columnwidth]{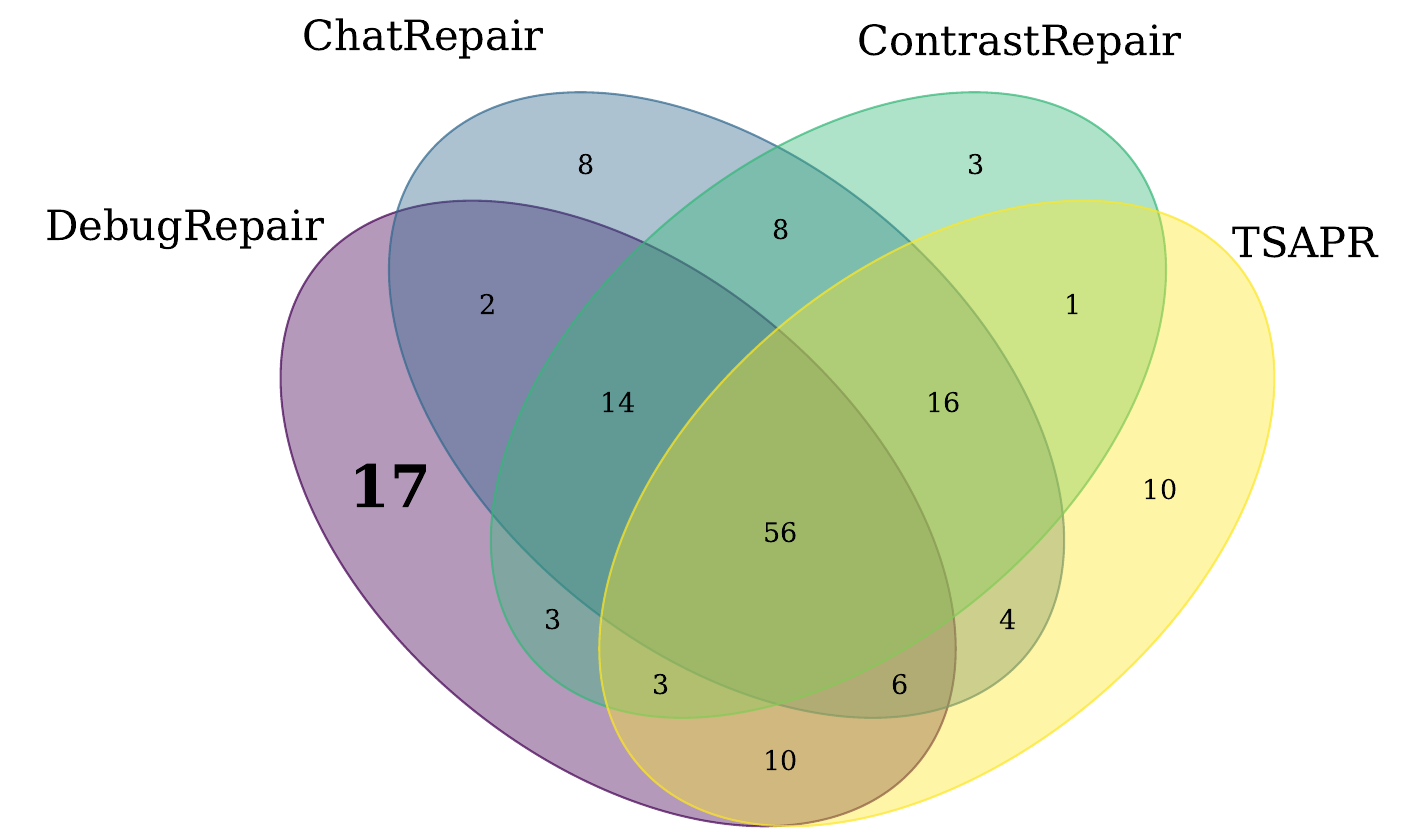}
        \label{venn_d4j12_2}
    }
    \subfigure[Venn on Defects4J-V2.0] {
        \includegraphics[width=0.4\columnwidth]{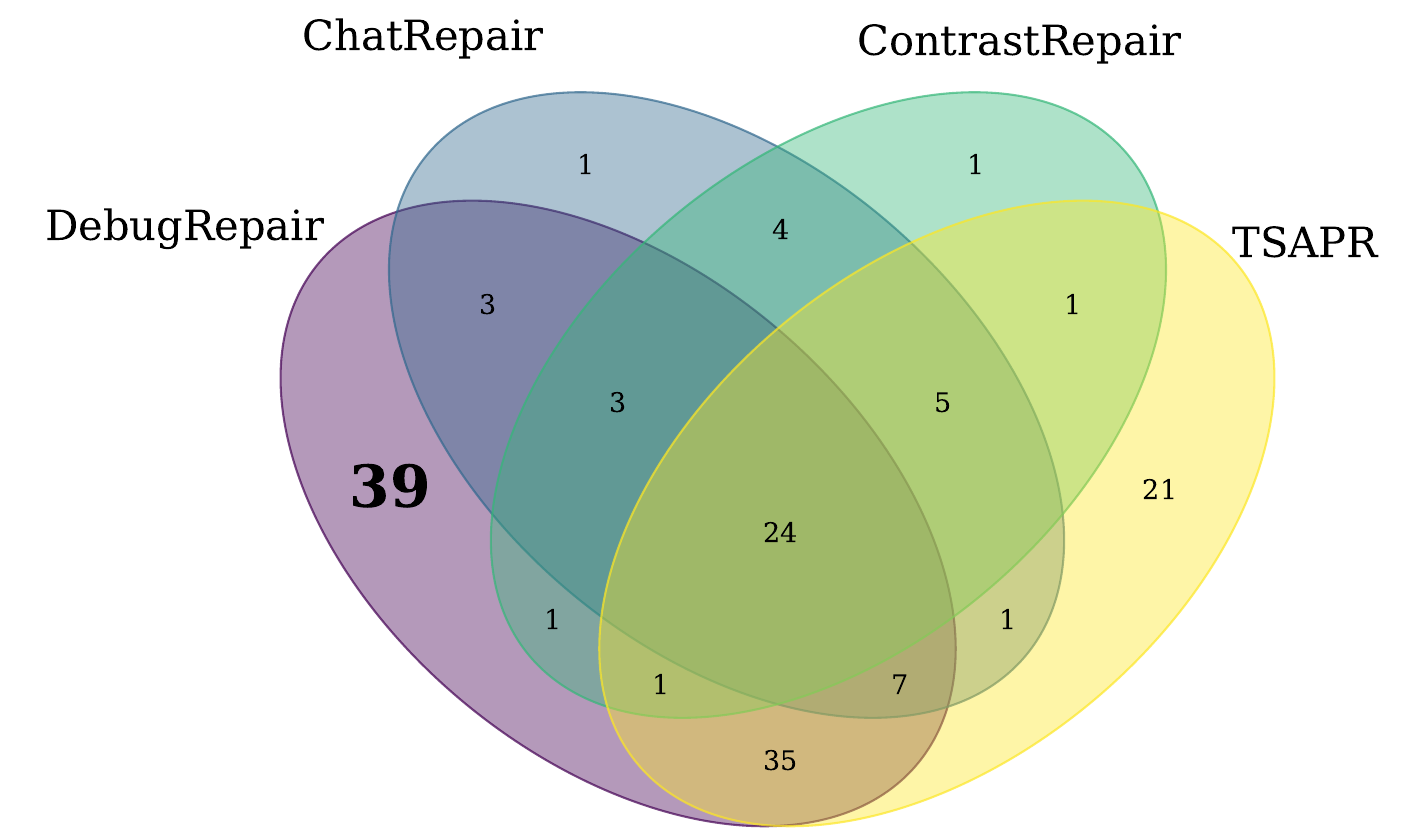}
        \label{venn_d4j20_2}
    }
    \vspace{-0.15in}
    \caption{Bug Fix Venn Diagram on Defects4J (DebugRepair, ChatRepair, ContrastRepair, TSAPR)}
    \Description{}
    \label{d4j_venn_2}
\end{figure}

\textbf{Unique Fix Analysis.} 
To better illustrate the distinct repair capabilities of DebugRepair, we conduct an overlap analysis of the correct fixes using Venn diagrams. Specifically, we present two groups of comparisons: one against three representative baselines (RepairAgent, ThinkRepair, and BaseChatGPT), and another against feedback-based baselines (ChatRepair, ContrastRepair, and TSAPR). For the first group, we compare DebugRepair with RepairAgent, ThinkRepair, and BaseChatGPT (excluding ReinFix, as its detailed results under the GPT-3.5 backbone are not publicly available). As shown in Figure \ref{venn_d4j12_1} and \ref{venn_d4j20_1}, 33 and 20 bugs on Defects4J-V1.2 and V2.0, respectively, can be successfully repaired by all these approaches. This overlap indicates that these approaches are highly effective and exhibit considerable similarity in their repair capabilities, largely because they are implemented on the same backbone model. Despite that, DebugRepair is still able to uniquely fix 27 and 22 bugs on Defects4J-V1.2 and V2.0, respectively, which are not fixed by any of these other approaches, and ranks first in unique fixes. For the second group, we analyze the overlap results among feedback-based baselines, including ChatRepair, ContrastRepair, and TSAPR. As shown in Figure \ref{venn_d4j12_2} and \ref{venn_d4j20_2}, 56 and 24 bugs on Defects4J-V1.2 and V2.0, respectively, can be successfully repaired by all these approaches, again indicating a considerable overlap in their repair capabilities. Meanwhile, DebugRepair fixes the largest number of bugs that remain unfixed by the others, contributing 17 and 39 unique fixes on V1.2 and V2.0, respectively. These results highlight the unique advantage of DebugRepair. In fact, since DebugRepair specifically focuses on patch refinement via dynamic debugging, it can be easily integrated into most existing APR tools, making it a strong complement to existing work.

\begin{figure}[htbp]
\centering
    \vspace{-0.1in} 
    \includegraphics[width=1.0\columnwidth]{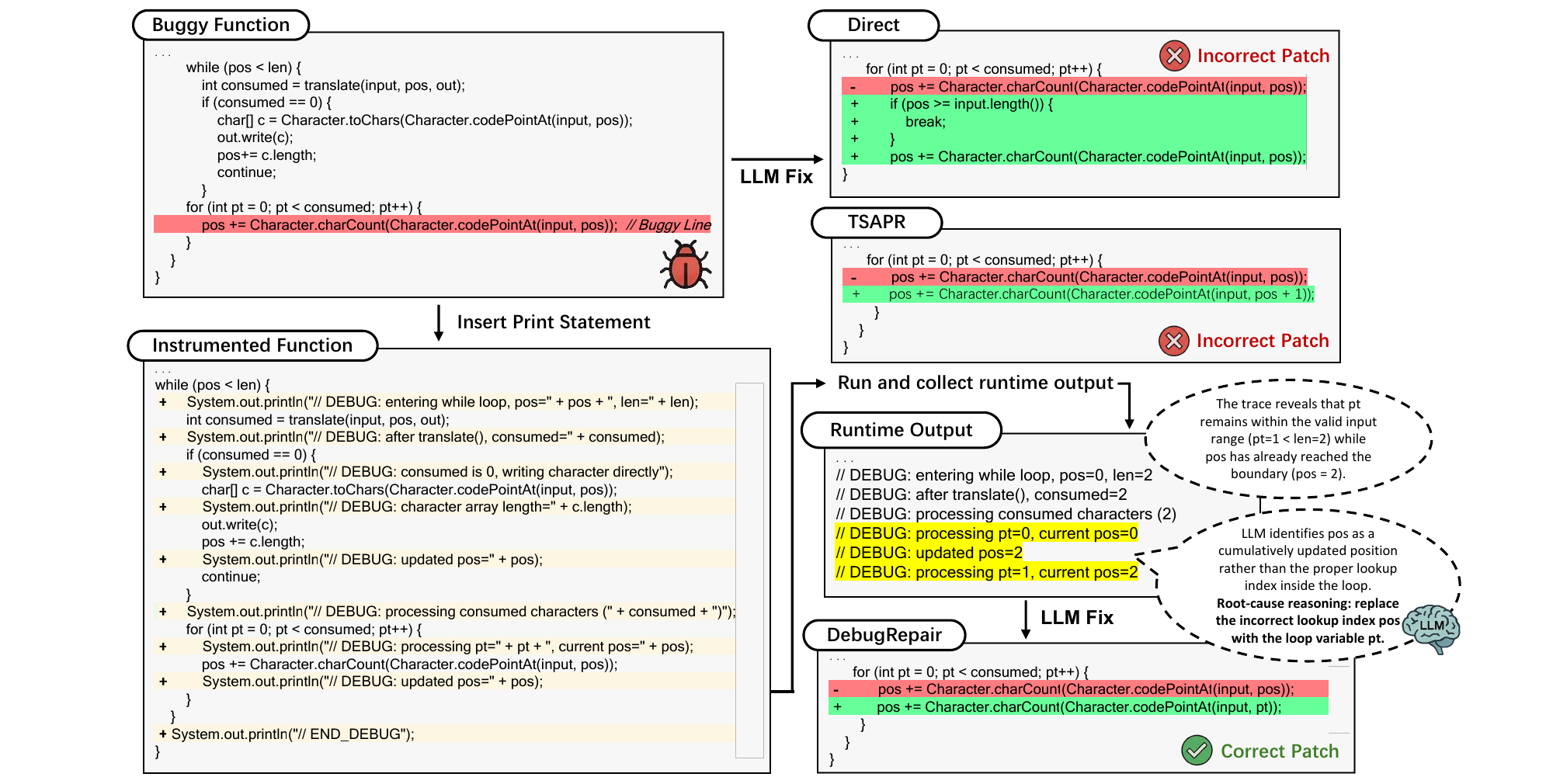}
    \vspace{-0.1in}  
    \caption{An illustration of the self-directed debugging process on the Lang-6 bug. }
    \Description{A workflow diagram showing the buggy code, the insertion of debug prints, the captured runtime output, and the final correct patch.}
    \label{fig:case_study}
    \vspace{-0.15in} 
\end{figure}

{
\textbf{Case Study.}
To further illustrate the effectiveness of DebugRepair and qualitatively analyze how runtime-state information affects LLM reasoning, we present a case study on the Lang-6 bug from Defects4J, as shown in Figure~\ref{fig:case_study}. DebugRepair successfully generates a correct patch for this bug, whereas all seven LLM-based baselines fail. This bug occurs in a character processing utility of Apache Commons Lang, where \texttt{pos} is repeatedly used as the index of \texttt{Character.\allowbreak codePointAt(input, pos)} while being cumulatively updated across loop iterations, eventually causing a \texttt{StringIndex\allowbreak OutOfBounds\allowbreak Exception}. Without runtime-state information, the direct repair attributes the failure to insufficient boundary protection and inserts a boundary check on \texttt{pos}; however, this symptom-oriented modification leaves the incorrect lookup index unchanged. Similarly, TSAPR uses execution feedback to guide Monte Carlo Tree Search (MCTS)-based patch exploration, but its feedback remains limited to outcome-level failure symptoms, leading to incorrect modifications such as changing the index from \texttt{pos} to \texttt{pos + 1}. In contrast, DebugRepair explicitly instruments the buggy function and exposes the evolution of intermediate states. The runtime trace shows that after the first iteration, \texttt{pos} increases from 0 to 2, while the next iteration starts with \texttt{pt=1} and \texttt{pos=2}. This evidence reveals that \texttt{pt} remains within the valid input range while \texttt{pos}, as a cumulatively updated position, has already reached the boundary. Based on this distinction, the LLM identifies that \texttt{pos} is incorrectly reused as the lookup index and replaces it with the loop variable \texttt{pt}, generating a patch consistent with the developer fix. Notably, unlike TSAPR, which searches over multiple patch candidates without success, DebugRepair only needs to iteratively debug a single candidate for a few rounds. This example highlights that intermediate runtime states help the LLM move beyond outcome-level failure symptoms to identify the underlying bug cause, thereby enabling more precise patch generation.
}

\begin{boxK}
\small \faIcon{pencil-alt} \textbf{Answer to RQ1:} 
{DebugRepair outperforms SOTA baselines. With GPT-3.5, it correctly fixes 111 and 113 bugs on Defects4J-V1.2 and V2.0, respectively, and all 40 bugs on QuixBugs. With DeepSeek-V3, DebugRepair fixes 295 Defects4J bugs, exceeding the strongest baseline by 59 correct fixes.}
\end{boxK}

\subsection{RQ2: Performance Across Repair Scenarios}
\label{RQ2: Performance Across Repair Scenarios}

To evaluate the robustness of DebugRepair, we investigate its performance across different dimensions of bug complexity: Single-Function (SF), Single-Hunk (SH), and Single-Line (SL) scenarios. Since the results from RQ1 demonstrate that LLM-based approaches achieve SOTA performance and outperform traditional and earlier learning-based approaches, we focus our comparative analysis in this section exclusively on these LLM-based baselines. 
Table \ref{tab:result_three_repair_scenarios} presents the comparative results on Defects4J and QuixBugs, where GPT-3.5 is integrated as the backbone LLM for each approach. 
{We also adopt DeepSeek-V3 as a backbone LLM for experiments in this RQ, with the corresponding results reported in Table~\ref{tab:result_deepseek} as part of RQ1.}

\begin{table}[h]
\scriptsize
\caption{Repair results of different repair scenarios for DebugRepair and baselines on Defects4J and QuixBugs (\# Correct/\# Plausible).}
\label{tab:result_three_repair_scenarios}
\resizebox{\columnwidth}{!}{
\begin{tabular}{@{}ccccccccccc@{}}
\toprule
\multirow{3}{*}{Category}        & \multirow{3}{*}{APR Approach} & \multicolumn{6}{c}{Defects4J}                                                                                                                                                                                                                                                                                                                                      & \multicolumn{3}{c}{QuixBugs}                                                                                                                                                      \\ \cmidrule(l){3-11} 
                                 &                             & \multicolumn{3}{c}{Defects4J-V1.2}                                                                                                                                                            & \multicolumn{3}{c}{Defects4J-V2.0}                                                                                                                                 & \multicolumn{2}{c}{Java}                                                                                              & Python                                                    \\ \cmidrule(l){3-11} 
                                 &                             & SF                                                             & SH                                                         & SL                                                              & SF                                                             & SH                                                              & SL                              & SF                                                        & SH                                                        & SF                                                        \\ \midrule
Basic                            & BaseChatGPT                 & 75/104                                                         & 56/74                                                      & 36/45                                                           & 67/100                                                         & 49/73                                                           & 31/44                           & 33/36                                                     & 33/36                                                     & 32/37                                                     \\ \midrule
\multirow{2}{*}{Retrieval-based} & RepairAgent                 & 80/-                                                           & 68/-                                                       & 49/-                                                            & 65/-                                                           & 62/-                                                            & 45/-                            & -                                                         & -                                                         & -                                                         \\
                                 & ReinFix                     & 104/-                                                          & 78/-                                                       & 53/-                                                            & \underline{109}/-                                & \textbf{85}/-                                  & 47/-                            & -                                                         & -                                                         & -                                                         \\ \midrule
Hybrid                           & ThinkRepair                 & 98/-                                                           & 78/-                                                       & 52/-                                                            & 107/-                                                          & 81/-                               & 47/-                            & 39/-                                                      & 36/-                                                      & \textbf{40}/-                            \\ \midrule
\multirow{4}{*}{Feedback-based}  & ChatRepair                  & 76/-                                                           & 79/-                          & \textbf{57}/-                                  & -                                                              & -                                                               & \underline{48}/-  & 39/-                                                      & \textbf{37}/-                            & \textbf{40}/-                            \\
                                 & ContrastRepair              & 75/\underline{120}                                                         & 69/\underline{102}                                                     & \underline{56}/60 & -                                                              & -                                                               & 40/-                            & \textbf{40}/-                            & -                                                         & \textbf{40}/-                            \\
                                 & TSAPR                       & \underline{108}/\textbf{146}    & \textbf{86}/\textbf{104} & \textbf{57}/\textbf{64}       & 93/\underline{134}                                                         & 77/\textbf{102}                                & 45/\textbf{55} & \textbf{40}/-                            & -                                                         & -                                                         \\
                                 & DebugRepair                 & \textbf{111}/\textbf{146} & \underline{82}/\underline{102}                                                      & 55/\underline{61}                                                           & \textbf{113}/\textbf{137} & \underline{83}/\underline{100} & \textbf{50}/\textbf{55}                           & \textbf{40}/\textbf{40} & \textbf{37}/\textbf{37} & \textbf{40}/\textbf{40} \\ \bottomrule
\end{tabular}
}
\begin{tablenotes}
\footnotesize
\item $^{\Phi}$ ``-'' indicates no results reported in the original work. In addition, the highest numbers of plausible and correct fixes are highlighted in \textbf{bold}, while the second-ranked ones are highlighted in \underline{underlines}.
\end{tablenotes}
\end{table}

As shown in Table \ref{tab:result_three_repair_scenarios}, under the GPT-3.5 backbone, DebugRepair demonstrates superiority in handling complex repair scenarios (i.e., SF and SH bugs), consistently performing at the SOTA level. In these complex settings, DebugRepair frequently outperforms the top-performing LLM-based approaches, such as ReinFix and TSAPR. For example, on Defects4J-V1.2, DebugRepair achieves the highest number of SF fixes (111), directly outperforming TSAPR (108) and ReinFix (104). On Defects4J-V2.0, DebugRepair further solidifies its lead in the SF scenario by fixing 113 bugs, surpassing ReinFix (109) and TSAPR (93). In the SH scenarios, DebugRepair maintains highly competitive performance, fixing 82 and 83 bugs on V1.2 and V2.0, respectively, closely rivaling the top baseline results in these categories (86 for TSAPR on V1.2, and 85 for ReinFix on V2.0). As a feedback-based approach, DebugRepair can be utilized by ReinFix for patch refinement owing to their orthogonal working mechanisms. As for TSAPR, although it follows the same paradigm of utilizing execution feedback as DebugRepair, ChatRepair, and ContrastRepair, its underlying design philosophy is different. Specifically, TSAPR concentrates on utilizing feedback to guide an MCTS for patch exploration. In contrast, approaches such as ChatRepair, ContrastRepair, and DebugRepair focus directly on the iterative refinement of a given candidate patch. Hence, the latter category is also orthogonal to TSAPR and can be integrated into it for further improvement during the patch refinement. 
Under this refinement-oriented philosophy, DebugRepair significantly outperforms ChatRepair and ContrastRepair, correctly fixing 111 SF and 82 SH bugs on Defects4J-V1.2, compared with 76 SF and 79 SH bugs for ChatRepair and 75 SF and 69 SH bugs for ContrastRepair.

This effectiveness in complex repairs is further corroborated by the results on the QuixBugs benchmark, where DebugRepair successfully fixes 100\% of the SF and SH bugs. 
{Furthermore, the experiments conducted with the DeepSeek-V3 backbone confirm that this advantage scales well. Powered by DeepSeek-V3, DebugRepair maintains strong performance in complex scenarios against competitive baselines such as ReinFix and TSAPR. To be specific, DebugRepair correctly fixes 139 and 156 SF bugs on Defects4J-V1.2 and V2.0, respectively, outperforming ReinFix (118 and 118) and TSAPR (108 and 116). A similar advantage is observed in the SH scenarios, where DebugRepair achieves 98 and 113 fixes, compared with 87 and 89 for TSAPR and 83 and 89 for ReinFix, respectively.}
Importantly, because DebugRepair specifically targets the patch refinement through dynamic execution states, its mechanism is highly orthogonal to existing APR tools, such as ReinFix and TSAPR. This orthogonality indicates a strong potential for integrating these approaches in the future to further boost repair performance.

In the simpler Single-Line (SL) repair scenarios, DebugRepair continues to demonstrate highly competitive and even leading performance. Under the GPT-3.5 backbone, DebugRepair attains 55 and 50 correct fixes for SL bugs on the Defects4J-V1.2 and V2.0 datasets, respectively. On Defects4J-V1.2, it performs neck-and-neck with other top approaches, falling only marginally behind ChatRepair and TSAPR (57 fixes). Notably, on Defects4J-V2.0, DebugRepair surpasses all evaluated baselines. 
{This strong performance is similarly observed when using the DeepSeek-V3 backbone. With this model, DebugRepair achieves 57 SL fixes on Defects4J-V1.2, which closely follows the best-performing approach, namely ContrastRepair's 60 fixes, and secures 61 SL fixes on Defects4J-V2.0 to significantly outperform the second-ranked ContrastRepair (48).}
The marginal variance observed on Defects4J-V1.2 compared to the top baselines is anticipated. SL bugs typically require straightforward pattern matching based on outcome-level failure symptoms. In these cases, inserting print statements and collecting execution traces can make the input context longer and slightly more difficult for the LLM to process. Despite this, DebugRepair still maintains a top-tier repair rate, and the robust capability is further evidenced on the QuixBugs benchmark, which consists predominantly of SL bugs. On this dataset, DebugRepair achieves a 100\% fix rate across both Java and Python versions.
Overall, these findings demonstrate that while the self-directed debugging strategy excels in complex, multi-statement scenarios, it remains a highly effective framework for simpler, line-level fixes.



\begin{boxK}
\small \faIcon{pencil-alt} \textbf{Answer to RQ2:} 
DebugRepair achieves SOTA performance on complex repairs while maintaining competitive performance on simpler bugs, demonstrating its effectiveness across different repair scenarios.
\end{boxK}

\subsection{RQ3: Generalizability Across Different LLMs}


{To evaluate the model-agnostic nature of DebugRepair, we implement DebugRepair across six distinct LLMs, encompassing general models, code models, and commercial models.} Table \ref{tab:result_llm} summarizes the performance improvements achieved by DebugRepair over the vanilla backbone models. 

\begin{table}[h]
\scriptsize
\caption{Comparison results between vanilla LLMs and DebugRepair on Defects4J (\# Correct/\# Plausible).}
\label{tab:result_llm} 
\begin{tabular}{@{}llccccccc@{}}
\toprule
                             &                                                        & \multicolumn{3}{c}{Defects4J-V1.2}                                             & \multicolumn{3}{c}{Defects4J-V2.0}                                            &                          \\ \cmidrule(lr){3-8}
\multirow{-2}{*}{Category}   & \multirow{-2}{*}{Model}                                & SF                       & SH                       & SL                       & SF                       & SH                       & SL                       & \multirow{-2}{*}{Total}  \\ \midrule
                             & Qwen2.5-7B                                             & 40/64                         &30/46                          &20/29                          & 46/73                         & 36/57                         & \textbf{23/35}                         &86/137                          \\
                             & Qwen2.5-7B (DebugRepair)       &\textbf{50/65}  &\textbf{39/49}  &\textbf{27/32}  &\textbf{57/75}  &\textbf{42/58}  &22/32  &\textbf{107/140}  \\
                             & Qwen2.5-32B                                            & 62/90                         &47/64                          &33/43                         &  62/88                        &  42/63                        &   23/34                       &124/178                          \\
\multirow{-4}{*}{General}    & Qwen2.5-32B (DebugRepair)      &\textbf{92/126}  &\textbf{66/89}  &\textbf{39/50}  &\textbf{103/121}  &\textbf{78/91}  &\textbf{38/48}  &\textbf{195/247}  \\ \midrule
                             & Qwen2.5-Coder-7B                                       & 58/88                         & 46/67                         &33/45                          & 53/82                         &   40/62                       &    27/36                      &111/170                          \\
\multirow{-2}{*}{Code}       & Qwen2.5-Coder-7B (DebugRepair) &\textbf{62/96}  &\textbf{54/80}  &\textbf{33/47}  &\textbf{79/91}  &\textbf{61/69}  &\textbf{34/38}  &\textbf{141/187}  \\ \midrule
                             & GPT-3.5                                                &75/104                          &56/74                          &36/45                          &67/100                          &49/73                          &31/44                          &142/204                          \\
                             & GPT-3.5 (DebugRepair)          &\textbf{111/146}  &\textbf{82/102}  &\textbf{55/61}  &\textbf{113/137}  &\textbf{83/100}  &\textbf{50/55}  &\textbf{224/283}  \\
                             & DeepSeek-V3                                            & 82/112                         & 59/76                         & 36/46                         & 73/91                         & 56/70                         & 31/40                         & 155/203                         \\
                             & DeepSeek-V3 (DebugRepair)      &\textbf{139/179}  &\textbf{98/122}  &\textbf{57/65} &\textbf{156/180}  &\textbf{113/129}  &\textbf{61/65}  &\textbf{295/359}  \\
                             & GPT-5.5                                                &112/166                          &67/92                          &40/54                          &157/213                          &102/128                          &61/76                          &269/379                          \\
\multirow{-6}{*}{Commercial} & GPT-5.5 (DebugRepair)          &\textbf{182/251}  &\textbf{109/139}  &\textbf{64/80}  &\textbf{177/223}  &\textbf{110/131}  &\textbf{68/77}  &\textbf{359/474}  \\ \bottomrule
\end{tabular}
\begin{tablenotes}
\footnotesize
\item $^{\Phi}$ Note that, for each backbone model, the highest numbers of plausible and correct fixes are highlighted in \textbf{bold}.
\end{tablenotes}
\vspace{-0.1in}
\end{table}

{As illustrated in Table \ref{tab:result_llm}, DebugRepair universally enhances the repair effectiveness of all evaluated LLMs, achieving an average performance improvement of 48.4\% over the vanilla backbone models.} Specifically, for general models, the integration of DebugRepair yields substantial improvements, increasing the number of correct fixes for Qwen2.5-7B and Qwen2.5-32B by 21 and 71, respectively. Similarly, for code models, Qwen2.5-Coder-7B achieves an increase of 30 correct fixes. {For commercial models, DebugRepair increases the total correct fixes from 142 to 224 for GPT-3.5, from 155 to 295 for DeepSeek-V3, and from 269 to 359 for GPT-5.5.} These results demonstrate that DebugRepair effectively enhances APR performance across different categories of LLMs.

A horizontal comparison across the evaluated models reveals clear performance hierarchies regarding base architectures and parameter sizes. First, commercial models consistently achieve the highest overall repair performance, followed by code models, and finally general models. These results are similar to the findings from previous studies \cite{xue2025classeval, xue2025new} on software engineering tasks. Specifically, at the same parameter scales, code models consistently outperform their general model counterparts; for example, the DebugRepair-enhanced Qwen2.5-Coder-7B resolves 141 bugs, significantly outperforming the general model Qwen2.5-7B and approaching the efficacy of the much larger vanilla GPT-3.5. 
Second, repair effectiveness scales predictably with model size within the same model family. For example, Qwen2.5-32B consistently outperforms Qwen2.5-7B in both the vanilla and DebugRepair-enhanced settings.


Further analysis of the absolute performance gains reveals several insights into how different models benefit from dynamic feedback. 
{Within the Qwen2.5 family, larger models appear to benefit more from DebugRepair, with Qwen2.5-32B gaining 71 additional correct fixes compared with 21 for Qwen2.5-7B. Moreover, DebugRepair shows a stronger improvement on the code-specialized Qwen2.5-Coder-7B (+30 fixes) than on the general Qwen2.5-7B (+21 fixes). Notably, the commercial models exhibit consistently larger absolute gains, with improvements of 82, 140, and 90 correct fixes for GPT-3.5, DeepSeek-V3, and GPT-5.5, respectively. Among them, DeepSeek-V3 achieves the largest improvement (+140 fixes). Meanwhile, even the substantially stronger GPT-5.5, which already fixes 269 bugs in the vanilla setting, further benefits from DebugRepair and reaches 359 correct fixes, yielding 90 additional fixes. These results demonstrate that DebugRepair provides substantial improvements across models with different capacities and specializations, including already strong commercial LLMs.}

\begin{boxK}
\small \faIcon{pencil-alt} \textbf{Answer to RQ3:} 
DebugRepair consistently enhances the repair capability of various LLMs, regardless of their size, type, or architecture. {On average, it improves the number of correct fixes by 48.4\% across the evaluated models.}
\end{boxK}

\subsection{RQ4: Ablation Study}




To understand the contribution of each component in DebugRepair, we conduct an ablation study by constructing five variants. (1) \textit{DebugRepair-w/o-Purification} uses the raw failing test without slicing, proceeding directly to the subsequent debugging process. {(2) \textit{DebugRepair-w/o-Debugging} disables the simulated debugging stage and relies solely on outcome-level failure symptoms for repair, while keeping all other components and repair settings unchanged.} (3) \textit{DebugRepair-w/o-Augmentation} returns the first plausible patch found without generating other variants. Additionally, to further analyze the instrumentation design within the simulated debugging stage, we construct two additional variants: (4) \textit{DebugRepair-wo-LLM Instrumentation}, which removes LLM-based instrumentation and uses only the rule-based instrumentation strategy; and (5) \textit{DebugRepair-wo-Rule Instrumentation}, which removes the rule-based fallback and skips cases where LLM-generated instrumentation fails. Table~\ref{tab:result_ablation} reports the results on the Defects4J dataset using GPT-3.5 as the backbone LLM.


\begin{table}[h]
\centering
\scriptsize
\caption{Ablation results of DebugRepair with different components and instrumentation strategies.}
\label{tab:result_ablation} 
\begin{threeparttable}
\setlength{\tabcolsep}{9pt}
\begin{tabular}{@{}lcccc@{}}
\toprule
Variant & \# Plausible & Plausible Drop & \# Correct & Correct Drop \\ 
\midrule
w/o-Purification            & 215 & $\downarrow$ 24.0\% & 164 & $\downarrow$ 26.8\% \\
w/o-Debugging               & 213 & $\downarrow$ 24.7\% & 165 & $\downarrow$ 26.3\% \\
w/o-Augmentation            & \textbf{283} & --                  & 173 & $\downarrow$ 22.8\% \\
\midrule
w/o-LLM-based Instrumentation     & 233 & $\downarrow$ 17.7\% & 189 & $\downarrow$ 15.6\% \\
w/o-Rule-based Instrumentation    & 227 & $\downarrow$ 19.8\% & 181 & $\downarrow$ 19.2\% \\
\midrule
DebugRepair                 & \textbf{283} & --                  & \textbf{224} & --                  \\ 
\bottomrule
\end{tabular}
\begin{tablenotes}
\footnotesize
\item $^{\Phi}$ Note that the highest numbers of plausible and correct fixes are highlighted in \textbf{bold}.
\end{tablenotes}
\end{threeparttable}
\vspace{-0.1in}
\end{table}


\textbf{Effectiveness of test purification.} Removing test semantic purification leads to a substantial performance decline, with the number of correct fixes dropping from 224 to 164, corresponding to a 26.8\% decrease.
This result confirms that real-world failing tests often contain irrelevant test scenarios that obscure the failure-triggering logic and bring about redundant runtime outputs. To quantitatively validate this, we further analyze the runtime output collected from the inserted print statements during the repair process. We observe that applying test purification reduces the average token count of the collected runtime output by 18.6\% compared to the unpurified setting. This reduction indicates that test purification effectively minimizes redundant runtime outputs, ensuring the concentration of LLMs during bug fixing. 



\textbf{Effectiveness of simulated debugging.} Disabling the simulated debugging mechanism also causes a substantial performance drop, reducing the number of plausible and correct fixes to 213 and 165, respectively, with a 24.7\% and 26.3\% decrease. This result highlights the importance of dynamic execution evidence. Without runtime traces, the framework has to rely only on static, outcome-level error information, which greatly limits the LLM’s ability to reason about intermediate program states and trace the root cause of complex bugs. The result shows that incorporating runtime trace information is essential for grounding patch generation in concrete execution behavior.


\textbf{Effectiveness of patch augmentation.} Removing the patch augmentation stage preserves the number of plausible patches (283) but reduces the number of correct fixes from 224 to 173, corresponding to a 22.8\% decrease. This gap directly reflects the common overfitting issue in APR: a patch may satisfy the available test suite while still being semantically incorrect. Patch augmentation alleviates this issue by generating and validating diverse semantic variants from the initial plausible patch, thereby increasing the chance of obtaining a truly correct fix.

\begin{figure}[htbp]
\centering
    \vspace{-0.1in} 
    \includegraphics[width=0.5\columnwidth]{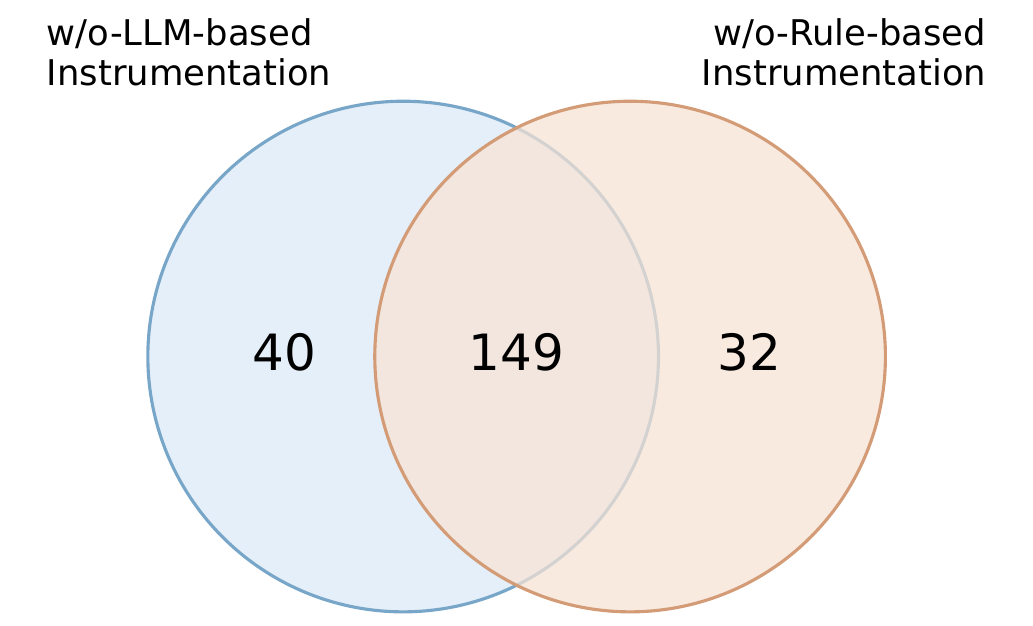}
    \vspace{-0.1in}  
    \caption{Overlap of correct fixes between the two single-strategy instrumentation variants. }
    \label{fig:venn_instrumentation}
    \vspace{-0.15in} 
\end{figure}

\textbf{Effectiveness of the hybrid instrumentation design.} 
We further study the internal design of the simulated debugging stage by separately ablating its two instrumentation strategies. When LLM-based instrumentation is removed, and only the rule-based strategy is retained, the number of correct fixes drops from 224 to 189. When the rule-based fallback is removed and failed LLM instrumentation cases are skipped, the number further drops to 181. Similar performance drops are also shown in plausible fixes.
{To further illustrate their complementarity, Figure~\ref{fig:venn_instrumentation} shows the overlap of correct fixes produced by the two single-strategy variants. Among the 189 and 181 bugs fixed by \textit{w/o-LLM-based Instrumentation} and \textit{w/o-Rule-based Instrumentation}, respectively, 149 are shared, while 40 and 32 are uniquely fixed by each variant.
The above results suggest that the two strategies are complementary. LLM-based instrumentation provides flexible, context-aware observation points, while the rule-based strategy improves robustness by guaranteeing trace collection when LLM-generated instrumentation is invalid or fails the consistency check. Moreover, the full DebugRepair fixes 224 bugs, further demonstrating the benefit of combining both strategies.}

\begin{boxK}
\small \faIcon{pencil-alt} \textbf{Answer to RQ4:} 
The effectiveness of DebugRepair comes from the joint contribution of its three core components and its hybrid instrumentation design. Removing any single module leads to a substantial drop in correct fixes (22.8\%--26.8\%), proving that all components are jointly essential. Further ablation within the simulated debugging stage shows that both LLM-based instrumentation and the rule-based fallback are necessary.

\end{boxK}

\subsection{RQ5: Generality to Unseen Bugs}

In light of prevailing concerns regarding potential data leakage from established benchmarks, such as Defects4J and QuixBugs, during the pre-training phase of LLMs, we conduct an additional evaluation to assess the effectiveness of DebugRepair on unseen datasets. It is worth noting that while benchmark memorization is a common threat to validity, our comparative analysis in previous RQs remains robust, given that the evaluated baselines utilize the same backbone models. 
{To further investigate whether the effectiveness of DebugRepair persists under reduced data leakage risk, we evaluate DebugRepair on both HumanEval-Java and RWB, which were released or collected after the training data cutoff of the backbone model (i.e., GPT-3.5).}
As we mentioned in Section \ref{RQ2: Performance Across Repair Scenarios}, although ChatRepair, ContrastRepair, and TSAPR all follow a paradigm of refining patches with feedback, the first two focus more on the refinement of a given patch, while the last one concentrates on searching for a high-quality patch. Thus, DebugRepair is orthogonal to TSAPR but matches the philosophy of the former. As such, we use ChatRepair and ContrastRepair as baselines for comparison in this RQ.
{For HumanEval-Java, we directly adopt their reported results from \cite{kong2025contrastrepair}. For RWB, we reproduce both baselines using the same GPT-3.5 backbone.
}


{
As illustrated in Table \ref{tab:result_humanjava}, DebugRepair achieves the best results among the evaluated tools on the HumanEval-Java dataset, successfully fixing 153 out of 163 total bugs. In comparison, ContrastRepair achieves 137 correct fixes, while ChatRepair achieves 130. DebugRepair therefore improves the correct repair rate by 11.68\% and 17.69\% over ContrastRepair and ChatRepair, respectively. 
On RWB, DebugRepair correctly fixes 21 out of 44 bugs on RWB-V1.0 and 14 out of 29 bugs on RWB-V2.0. In comparison, ContrastRepair fixes 13 and 11 bugs, while ChatRepair fixes 9 and 8 bugs on the two versions, respectively. Overall, DebugRepair correctly fixes 35 out of 73 RWB bugs, outperforming ContrastRepair and ChatRepair, which fix 24 and 17 bugs, respectively.
Furthermore, DebugRepair achieves these results with a substantially smaller patch size than the baseline approaches.
Taken together, these results show that DebugRepair maintains strong repair effectiveness on both HumanEval-Java and RWB, providing further evidence of DebugRepair's generality to unseen bugs under reduced data leakage risk.
}

\begin{table}[h]
\centering
\scriptsize
\setlength{\tabcolsep}{14pt}
\caption{Repair results for DebugRepair and baselines on HumanEval-Java and RWB.}
\label{tab:result_humanjava} 
\begin{threeparttable}
\begin{tabular}{@{}lcccc@{}}
\toprule
Approach & Patch Size & HumanEval-Java & RWB-V1.0 & RWB-V2.0 \\ 
& & \# Correct & \# Correct & \# Correct \\ 
\midrule
ChatRepair     
& 500 
& 130 
& 9 
& 8 \\

ContrastRepair 
& 160 
& 137 
& 13 
& 11 \\

DebugRepair    
& 32 
& \textbf{153} 
& \textbf{21} 
& \textbf{14} \\ 
\bottomrule
\end{tabular}
\begin{tablenotes}
\footnotesize
\item $^{\Phi}$ Note that the highest numbers of correct fixes are highlighted in \textbf{bold}.
\end{tablenotes}
\end{threeparttable}
\vspace{-0.1in}
\end{table}

\begin{boxK}
\small \faIcon{pencil-alt} \textbf{Answer to RQ5:} 
{DebugRepair maintains high effectiveness on unseen bugs. Its consistent advantages on both HumanEval-Java and RWB benchmarks further demonstrate that its performance stems from the debugging-driven reasoning process rather than data memorization.}
\end{boxK}

\section{Discussion}


\subsection{Robustness under Realistic Fault Localization}
\label{fl}

{
In the preceding RQs, we evaluate DebugRepair under the assumption of perfect FL, which is widely adopted in recent LLM-based APR studies to provide a controlled setting for comparing patch generation capabilities. Nevertheless, in practical APR scenarios, fault locations are typically produced by automated FL techniques and can be imprecise. To assess the practical applicability of DebugRepair, we further evaluate its performance under a more realistic FL setting on Defects4J.

Following prior studies \cite{jiang2018shaping, li2020dlfix, liu2019tbar, wen2018context, ye2024iter}, we employ Spectrum-Based Fault Localization (SBFL) using GZoltar \cite{campos2012gzoltar} with the Ochiai \cite{abreu2007accuracy} formula to compute the suspiciousness score of each statement. Since DebugRepair performs repair at the function level, we retain the buggy function as the repair scope and replace the oracle statement-level FL used in the main experiments with the Top-1 suspicious statement identified by SBFL within that function. This suspicious statement is then provided to DebugRepair as the FL for subsequent instrumentation and repair. We conduct the experiments with both GPT-3.5 and DeepSeek-V3, while keeping all other repair configurations identical to those in the main experiments.

As shown in Table \ref{tab:result_realistic_fl}, under realistic FL, DebugRepair correctly fixes 164 and 225 bugs with GPT-3.5 and DeepSeek-V3, respectively, retaining 73.2\% and 76.3\% of the correct fixes achieved under perfect FL. These results indicate that DebugRepair remains effective even when the provided FL is imprecise. This robustness benefits from its debugging-oriented design, where the localized statement serves only as an initial repair hint, while the subsequent instrumentation further collects intermediate runtime states across the buggy function to guide patch refinement. Therefore, DebugRepair does not rely solely on the exact FL and can partially tolerate localization noise, demonstrating its practical applicability under more realistic APR settings.
}

\begin{table}[h]
\centering
\scriptsize
\setlength{\tabcolsep}{14pt}
\caption{Repair performance of DebugRepair under realistic fault localization settings on Defects4J.}
\label{tab:result_realistic_fl} 
\begin{threeparttable}
\begin{tabular}{@{}lccc@{}}
\toprule
Backbone Model 
& Perfect FL (\# Correct/\# Plausible) 
& Realistic FL (\# Correct/\# Plausible) 
& Correct Fix Retention \\ \midrule
GPT-3.5     
& 224/283 
& 164/231 
& 73.2\% \\
DeepSeek-V3 
& 295/359 
& 225/306 
& 76.3\% \\ \bottomrule
\end{tabular}
\begin{tablenotes}
\footnotesize
\item $^{\Phi}$ Correct Fix Retention denotes the proportion of correct fixes retained under realistic FL compared with perfect FL.
\end{tablenotes}
\end{threeparttable}
\vspace{-0.1in}
\end{table}

\subsection{Evaluation on the Recent Defects4J-V3.0 Benchmark}
\label{sec:d4j3}

{
To further evaluate DebugRepair on a more recent benchmark version, we conduct an additional experiment on Defects4J-V3.0. Following the repair scenarios defined in Section~\ref{bench}, we identify 524 SF bugs, including 305 SH and 176 SL bugs. Considering the experimental cost, we compare DebugRepair with ReinFix, which is the strongest baseline on Defects4J under both GPT-3.5 and DeepSeek-V3 in our main experiments. We evaluate both approaches using these two backbone LLMs, with the results reported in Table~\ref{tab:d4j3}.

As shown in Table~\ref{tab:d4j3}, with GPT-3.5, DebugRepair correctly fixes 134 SF and 92 SH bugs, outperforming ReinFix with 122 and 87 fixes, respectively. For SL bugs, the two approaches perform comparably, with 63 fixes for DebugRepair and 64 for ReinFix. With DeepSeek-V3, DebugRepair consistently outperforms ReinFix across all three scenarios, correctly fixing 267, 170, and 104 SF, SH, and SL bugs, compared with 240, 155, and 103 for ReinFix, respectively. These results are consistent with our findings in RQ2. Overall, this additional evaluation further demonstrates the effectiveness of DebugRepair on the more recent Defects4J-V3.0 benchmark.
}

\begin{table}[h]
\scriptsize
\setlength{\tabcolsep}{23pt}
\caption{Repair results of DebugRepair and ReinFix on Defects4J-V3.0 (\# Correct/\# Plausible).}
\label{tab:d4j3}
\centering
\begin{threeparttable}
\begin{tabular}{@{}llccc@{}}
\toprule
Backbone Model & APR Approach & SF (524) & SH (305) & SL (176) \\ 
\midrule

\multirow{2}{*}{GPT-3.5}
& ReinFix
& 122/166
& 87/112
& \textbf{64}/\textbf{80} \\

& DebugRepair
& \textbf{134}/\textbf{187}
& \textbf{92}/\textbf{119}
& 63/79 \\

\midrule

\multirow{2}{*}{DeepSeek-V3}
& ReinFix
& 240/324
& 155/199
& 103/124 \\

& DebugRepair
& \textbf{267}/\textbf{390}
& \textbf{170}/\textbf{238}
& \textbf{104}/\textbf{140} \\

\bottomrule
\end{tabular}

\begin{tablenotes}
\footnotesize
\item $^{\Phi}$ The highest numbers of plausible and correct fixes under each backbone model are highlighted in \textbf{bold}.
\end{tablenotes}
\end{threeparttable}
\end{table}

\subsection{Impact of Hyper-parameters}

\begin{figure}[htbp]
\centering
    \subfigure[Repair performance under different $N_{session}$ and $K_{round}$ settings.]{
        \includegraphics[width=1.0\columnwidth]{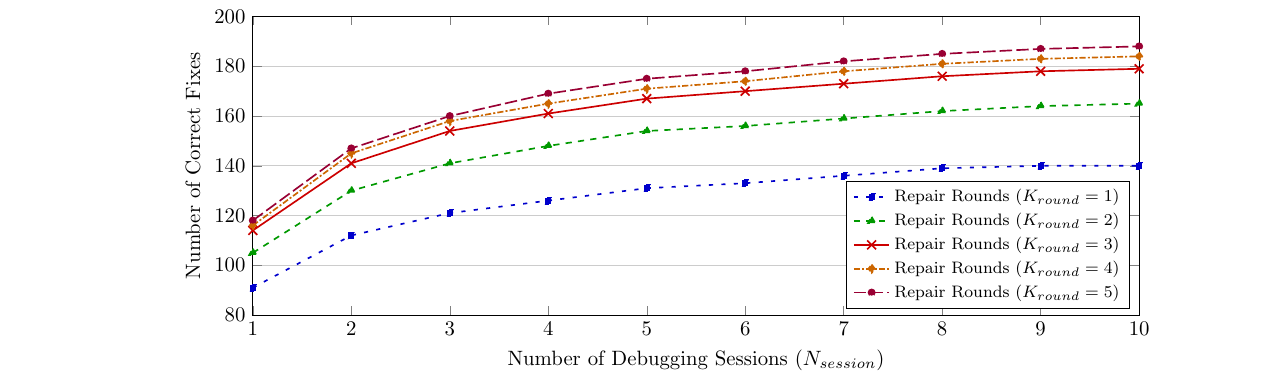}
        \label{mn}
    }
    \subfigure[Repair performance under different augmentation budgets.]{
        \includegraphics[width=1.0\columnwidth]{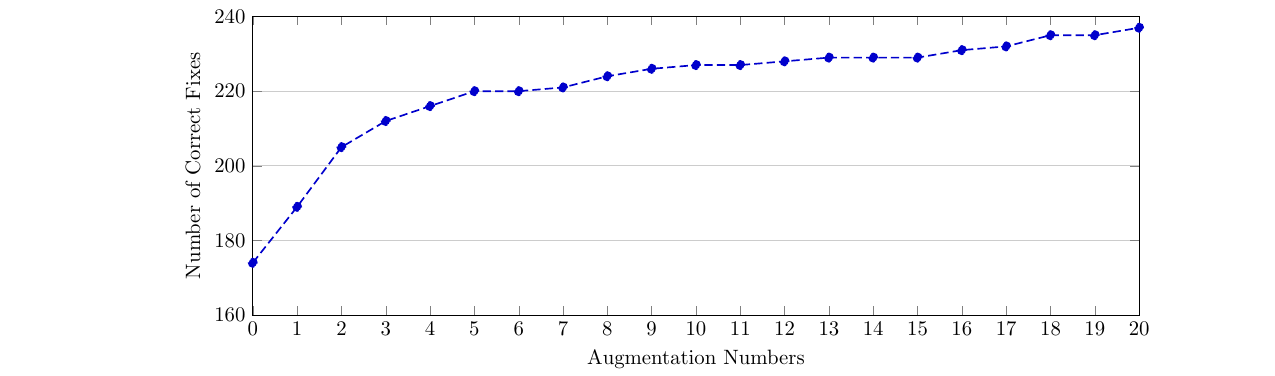}
        \label{k}
    }
    \vspace{-0.15in}
    \caption{Impact of hyper-parameter settings on the repair performance of DebugRepair.}
    \Description{}
    \label{hyper}
\end{figure}

We analyze the impact of DebugRepair’s hyper-parameters from two perspectives. We first study the interaction between the number of debugging sessions ($N_{session}$) and the number of repair rounds per session ($K_{round}$) without patch augmentation, so as to determine a proper repair budget for the main debugging process. We then fix this budget and investigate the effect of the augmentation number in the subsequent patch augmentation stage. As shown in Figure \ref{mn}, the number of correct fixes generally increases with both $N_{session}$ and $K_{round}$, indicating that enlarging either the cross-session re-debugging budget or the within-session iterative refinement budget can improve repair effectiveness.

A closer look at the trend reveals two important observations. First, increasing $K_{round}$ from 1 to 2 brings a particularly clear improvement, which shows that feedback-driven iterative refinement is highly beneficial. Second, as $N_{session}$ increases, performance rises rapidly in the early stage and then gradually saturates. This pattern suggests that restarting the repair process with newly collected debugging evidence is highly effective. At the same time, the later-stage gains become marginal, showing diminishing returns from continuously increasing the repair budget.

After fixing the main repair budget, we further study the augmentation number. As shown in Figure \ref{k}, increasing the number of augmentation queries consistently improves the number of correct fixes, confirming that patch augmentation is effective in transforming an initial plausible patch into more semantically correct variants. However, the improvement also becomes less pronounced when the augmentation budget grows larger, again indicating diminishing returns. Considering both repair effectiveness and computational overhead, we set $N_{session}=6$ and $K_{round}=4$ for the debugging-driven repair stage, and use 8 additional queries for patch augmentation. This yields the default configuration of $6 \times 4 + 8$, which provides a favorable trade-off between effectiveness and cost under a limited budget.

\subsection{Cost Analysis}
To comprehensively evaluate the practicality of the proposed framework, we analyze the cost differences between DebugRepair and existing APR tools in terms of patch size, token consumption, and monetary cost. We select ChatRepair \cite{xia2024automated}, RepairAgent \cite{bouzenia2024repairagent}, TSAPR \cite{hu2025tsapr}, and ReinFix \cite{zhang2025repair} as baselines, as other LLM-based baselines do not report cost-related statistics for direct comparison. For a fair comparison, all approaches use GPT-3.5 as the backbone model, and the cost metrics are evaluated on the Defects4J dataset.

Table \ref{tab:cost_comparison} presents the detailed cost comparison. With the patch size set to 32, which is tied for the smallest among all baselines alongside TSAPR, DebugRepair demonstrates a significant advantage in token efficiency. It consumes an average of only 38,000 tokens per bug, which is merely 18.1\% of the 210,000 tokens reported by ChatRepair, 14.1\% of the 270,000 tokens by RepairAgent, and less than the 40,000 tokens required by TSAPR. 
This token efficiency over TSAPR primarily stems from two factors: first, TSAPR incurs additional token overhead by invoking LLMs for intermediate reasoning (e.g., Chain-of-Thought) and patch evaluation (e.g., LLM-as-Judge), whereas DebugRepair limits LLM generation to instrumentation and patch generation; second, DebugRepair's higher repair success rate effectively reduces the average token consumption per fixed bug.

In terms of monetary cost, DebugRepair achieves the lowest cost at \$0.036 per bug. When compared to the recent SOTA baselines TSAPR and ReinFix (both costing \$0.06 per bug), DebugRepair achieves a substantial cost reduction of 40.0\%. Moreover, this cost is only 25.7\% of the \$0.14 expenditure reported by both ChatRepair (today's pricing) and RepairAgent. Crucially, DebugRepair achieves this superior cost-efficiency while simultaneously delivering a higher number of correct fixes, demonstrating the ability of the proposed framework to yield SOTA performance at a significantly lower budget.

{Beyond token and monetary costs, we further measure the end-to-end runtime of DebugRepair. On Defects4J with GPT-3.5, DebugRepair takes 19.92 minutes on average to successfully repair a bug. We further break down the runtime into four major categories: instrumentation preparation (e.g., instrumentation generation, consistency checking, and instrumented-code compilation) accounts for 18.35\%; instrumented execution (e.g., benchmark checkout, project build, and instrumented test execution) accounts for 3.62\%; runtime-trace processing (e.g., trace parsing and storage) accounts for only 0.07\%; and repair and validation (e.g., repair-model inference and patch validation) accounts for the remaining 77.96\%. These results show that the additional debugging-related operations do not dominate the overall runtime, with most of the cost arising from the repair process itself, indicating a reasonable trade-off between repair effectiveness and efficiency.}

\begin{table}[h]
\scriptsize
\setlength{\tabcolsep}{38pt}
\caption{Cost comparison between DebugRepair and existing APR approaches on Defects4J.}
\vspace{-0.1in}
\label{tab:cost_comparison}
\centering
\begin{threeparttable}
\begin{tabular}{@{}lccc@{}}
\toprule
Approach & Patch/Bug & Token/Bug & Money/Bug \\ \midrule
ChatRepair (2024) \cite{xia2024automated}      & 500       & 210,000   & \$0.42    \\
ChatRepair (today’s price)      & 500       & 210,000   & \$0.14    \\
RepairAgent (2024) \cite{bouzenia2024repairagent}     & 117       & 270,000   & \$0.14    \\
TSAPR (2025) \cite{hu2025tsapr} & \textbf{32}        & 40,000    & \$0.06    \\
ReinFix (2026) \cite{zhang2025repair}       & 45         & -         & \$0.06    \\
DebugRepair (Ours)                & \textbf{32}        & \textbf{38,000}    & \textbf{\$0.036}   \\ \bottomrule
\end{tabular}
\begin{tablenotes}
\footnotesize
\item $^{\Phi}$ ``-'' indicates no results reported in the original work, while the lowest cost on patch/token/money consumption per bug is highlighted in \textbf{bold}.
\end{tablenotes}
\end{threeparttable}
\vspace{-0.2in}
\end{table}

\section{Threats to Validity}
\textbf{Threats to construct validity.} 
The primary concern lies in the subjectivity of evaluating the correctness of generated patches. While automated testing can identify plausible patches that pass all test cases, verifying the semantic correctness of these patches requires manual inspection, which may introduce potential human bias. 
{To mitigate this threat, the two primary reviewers, each with 3–5 years of software development experience, independently assess the plausible patches while being blinded to the identity of the approach that generated them. The overall Cohen’s kappa coefficient \cite{wan2015kappa} between their initial assessments is 0.73, indicating substantial inter-rater agreement and supporting the reliability of our assessment process. Any remaining disagreements are further resolved through adjudication by a third independent researcher.} Furthermore, we have open-sourced all generated patches for public assessment \cite{146854hg69:online}.
{Another concern lies in monetary cost estimation. For comparability, all compared approaches use GPT-3.5 via the official OpenAI API, avoiding model- and provider-level differences; DebugRepair's cost is estimated using the current official pricing. Therefore, the reported monetary costs should be regarded as approximate and pricing-dependent estimates.}

\textbf{Threats to internal validity.}
A primary threat arises from potential data leakage in LLM-based evaluation. Since the backbone models are pretrained on large-scale public code corpora, they may have been exposed to repair patterns or corresponding fixes during training. 
{To mitigate this risk, we additionally evaluate DebugRepair on HumanEval-Java and RWB. HumanEval-Java was released after the training data cutoff of GPT-3.5, while both RWB versions contain bugs collected from bug-fixing commits after this cutoff. 
The consistent performance gains observed on these temporally separated benchmarks provide further evidence that the improvements of DebugRepair are not mainly attributable to memorization of pre-training data.}
{Another threat concerns potential semantic changes introduced by source-code instrumentation. Although DebugRepair mitigates this risk through prompt constraints, normalized-code equivalence and compilation checks, and deterministic rule-based fallback, semantic equivalence cannot be formally guaranteed. Empirically, only 10/483 and 15/483 Defects4J bugs show inconsistent test behavior after instrumentation under GPT-3.5 and DeepSeek-V3, respectively. In addition, manual inspection finds 192/200 sampled instrumentations to preserve the original semantics. These results suggest that such inconsistencies are rare, though they remain a residual threat to internal validity.}

\textbf{Threats to external validity.} 
External validity concerns the generalizability of the proposed approach across different datasets and programming languages (PLs). {To mitigate this threat, we evaluate DebugRepair on four widely used benchmarks, including Defects4J, QuixBugs, HumanEval-Java, and RWB, which differ in bug characteristics and complexity.} Notably, we intentionally exclude issue-resolving benchmarks such as SWE-bench \cite{jimenez2023swe} from our evaluation because they restrict APR tools from accessing explicit test cases as the default experimental setting, thereby rendering DebugRepair inapplicable to such evaluations.
Besides, our evaluation further spans both Java and Python to reduce potential PL-specific bias. 
{Although the Python evaluation is currently limited to QuixBugs, the consistent results on its Java and Python versions provide encouraging evidence that the core workflow of DebugRepair can be transferred across PLs with PL-specific adaptations. Further evaluation on larger real-world Python projects remains future work.}
{Another limitation concerns fault localization. Although Section \ref{fl} evaluates DebugRepair with realistic statement-level FL, the buggy function is still assumed to be known. Future work will integrate function-level FL into DebugRepair toward a more end-to-end repair process.}
{Finally, the current rule-based instrumentation fallback focuses on common AST node types and debugging scenarios rather than exhaustively covering all language constructs. Future work will further extend the rule-based mechanism with additional node-specific instrumentation strategies.}

\section{Related Work}

Automated Program Repair (APR) aims to automatically generate patches that fix software bugs, thereby reducing developers' manual debugging effort \cite{gazzola2018automatic, le2019automated, zhang2023survey}. Early APR studies can be broadly categorized into template-based \cite{martinez2016astor, hua2018sketchfix, ghanbari2019practical, liu2019avatar}, heuristic-based \cite{le2011genprog, jiang2018shaping, le2016history, wen2018context}, and constraint-based \cite{long2015staged, le2017s3, mechtaev2016angelix, gao2021beyond} paradigms.
For example, Liu et al. \cite{liu2019tbar} propose TBar, which applies a set of recurrently-used fix patterns to resolve software bugs. Gao et al. \cite{gao2021beyond} propose ExtractFix, which synthesizes vulnerability patches guided by crash constraints extracted via symbolic execution.
Although effective in specific settings, these approaches are often limited by limited patch diversity and limited generality for semantically complex bugs.

With advances in deep learning, APR shifted from rule-intensive repair to data-driven patch generation. 
Existing learning-based approaches mainly include NMT-based approaches \cite{chen2019sequencer, jiang2021cure, li2020dlfix, li2022dear}, which formulate repair as a code-to-code translation task, and PLM-based approaches \cite{xia2023plastic, wang2023rap, xia2023automated, xia2022less}, which leverage large-scale pre-training to improve contextual modeling and repair capability.
For example, Ye et al. \cite{ye2022selfapr} propose SelfAPR, a self-supervised training approach that captures fault-specific knowledge by encoding execution diagnostics. Additionally, Xia et al. \cite{xia2023plastic} introduce FitRepair, a PLM-based approach that uses fine-tuning and prompting to extract fix ingredients.
Despite their improved generality over traditional APR, these approaches still rely heavily on historical fixes and often struggle to reason about failure-specific execution semantics.

More recently, Large Language Models (LLMs) have substantially advanced APR due to their strong capabilities in code understanding and generation.
{Existing LLM-based APR techniques can be broadly categorized, according to their primary repair guidance mechanisms, into three paradigms: retrieval-based \cite{bouzenia2024repairagent, zhang2025repair}, feedback-based \cite{xia2024automated, kong2025contrastrepair}, and hybrid approaches \cite{yin2024thinkrepair}.} (1) Retrieval-based approaches, such as RepairAgent \cite{bouzenia2024repairagent} and ReinFix \cite{zhang2025repair}, enrich repair prompts with external code fragments, semantic ingredients, or historical fixes. (2) Feedback-based approaches leverage execution feedback to improve repair quality, but differ in how it is used: 
ChatRepair \cite{xia2024automated} and ContrastRepair \cite{kong2025contrastrepair} directly refine candidate patches through execution feedback, whereas TSAPR \cite{hu2025tsapr} incorporates execution feedback into a Monte Carlo Tree Search (MCTS) process to guide patch exploration.
(3) Hybrid approaches combine multiple repair signals to improve LLM-based patch generation. For example, ThinkRepair \cite{yin2024thinkrepair} adopts a two-phase framework that first collects chains of thought and verified fixes to build a knowledge pool, and then selects few-shot examples from this pool to guide repairs, optionally incorporating test-failure feedback for further refinement.
{Beyond the above repair guidance mechanisms, some recent approaches, such as RepairAgent \cite{bouzenia2024repairagent} and ReinFix \cite{zhang2025repair}, also employ agentic architectures. Since our taxonomy focuses on the primary source of repair guidance, we treat agenticity as an orthogonal design dimension rather than a separate repair paradigm.}

Despite the progress of LLM-based APR, a common limitation remains across these paradigms: 
Most approaches rely on source code and outcome-level failure symptoms, i.e., stack traces. Although useful for identifying the manifestation of a bug, these signals are often too coarse to expose the intermediate runtime states that directly reveal its root cause. As a result, LLMs may generate plausible yet semantically incorrect patches that only mask symptoms. We also found concurrent work, namely InspectCoder \cite{wang2026inspectcoder}, which follows the debugging philosophy as well, using natural-language task descriptions as explicit specifications. However, such an experimental setting (i.e., being equipped with task descriptions) differs from mainstream real-world APR studies, making it unfair to include it as a comparison.
In contrast, DebugRepair integrates test semantic purification, simulated instrumentation, and debugging-driven conversational repair, allowing LLMs to reason over intermediate runtime states and reduce the noise in failure-triggering test contexts, thereby providing a more failure-specific foundation for LLM-based APR.

\section{Conclusion}

In this paper, we propose DebugRepair, a novel LLM-based APR framework that enhances repair effectiveness via self-directed debugging. Unlike existing feedback-based APR approaches that mainly rely on outcome-level failure symptoms, DebugRepair equips LLMs with runtime-state evidence through three components, namely test semantic purification, simulated instrumentation, and debugging-driven conversational repair. 
{Experimental results on Defects4J, QuixBugs, HumanEval-Java, and RWB show that DebugRepair consistently achieves strong repair effectiveness, while ablation studies further confirm the effectiveness of all components in the framework.}

\vspace{-0.5em}

\section*{Acknowledgments}

This work was partially supported by the National Natural Science Foundation of China (Grant Nos. 62502283 and U24B20149), the Natural Science Foundation of Shandong Province (Grant No. ZR2024QF093), the Young Talent of Lifting Engineering for Science and Technology in Shandong, China (Grant No. SDAST2025QTB031).

\bibliographystyle{ACM-Reference-Format}
\AtEndEnvironment{thebibliography}{\enlargethispage{2\baselineskip}}
\bibliography{reference}

\end{document}